\documentclass[10pt,aps,prc,amsmath,amssymb,twocolumn,floatfix,nofootinbib,showpacs,noeprint,superscriptaddress]{revtex4-1}

\usepackage{graphicx}
\usepackage{color}
\usepackage{graphicx}
\usepackage{amsmath,amssymb,bm,amsfonts}
\usepackage{hyperref}

\graphicspath{{figures/}}

\definecolor{darkred}{rgb}{0.75, 0.0, 0.0}

\newcommand{\beq}{\begin{equation}}
\newcommand{\eeq}{\end{equation}}
\newcommand{\bea}{\begin{eqnarray}}
\newcommand{\eea}{\end{eqnarray}}

\newcommand{\hw}{\ensuremath{\hbar\Omega}}

\newcommand{\kinf}{\ensuremath{k_{\infty}}}
\newcommand{\kappainf}{\ensuremath{\kappa_{\infty}}}
\newcommand{\Einf}{\ensuremath{E_{\infty}}}

\newcommand{\rsqav}{\langle r^2 \rangle}
\newcommand{\rsqinf}{\ensuremath{\rsqav_{\infty}}}

\newcommand{\ANC}{\gamma_\infty}
\newcommand{\Jost}{\mathfrak{f}}

\newcommand{\LNLO}{\text{L-NLO}}

\begin{document}

\begin{titlepage}
  \title{Systematic expansion for infrared oscillator basis extrapolations}

  \author{R.J.\ Furnstahl} \email{furnstahl.1@osu.edu}
  \affiliation{Department of Physics, The Ohio State University,
    Columbus, OH 43210}

  \author{S.N.\ More} \email{more.13@osu.edu} \affiliation{Department
    of Physics, The Ohio State University, Columbus, OH 43210}

  \author{T.~Papenbrock} \email{tpapenbr@utk.edu}
  \affiliation{Department of Physics and Astronomy, University of
    Tennessee, Knoxville, Tennessee 37996, USA} \affiliation{Physics
    Division, Oak Ridge National Laboratory, Oak Ridge, Tennessee
    37831, USA}

\date{\today}

\begin{abstract}
  Recent work has demonstrated that the infrared effects of harmonic
  oscillator basis truncations are well approximated by a
  partial-wave Dirichlet boundary condition at a properly identified
  radius $L$.  This led to formulas for extrapolating the
  corresponding energy $E_L$ and other observables to infinite $L$ and
  thus infinite basis size.  Here we reconsider the energy for a
  two-body system with a Dirichlet boundary condition at $L$ to
  identify and test a consistent and systematic expansion for $E_L$
  that depends only on observables.  We also generalize the energy
  extrapolation formula to nonzero angular momentum, and apply it to
  the deuteron.  Formulas given previously for extrapolating the radius
  are derived in detail.
\end{abstract}

\smallskip
\maketitle

\end{titlepage}

\section{Introduction}

The use of finite harmonic oscillator (HO) model spaces in nuclear
structure calculations effectively imposes both infrared (IR) and
ultraviolet (UV) momentum cutoffs~\cite{Stetcu:2006ey,
  hagen:2010gd,jurgenson:2010wy,Coon:2012ab,Furnstahl:2012qg}.
Computational limits often require that the HO basis be truncated
before observables are fully converged, which has led to various
phenomenological schemes to extrapolate energies to infinite basis
size~\cite{Hagen:2007ew,Bogner:2007rx,Forssen:2008qp,Maris:2008ax,Roth:2009cw}.
More systematic development of extrapolation formulas is possible by
considering the IR and UV cutoffs explicitly, as first illustrated in
Ref.~\cite{Coon:2012ab}. A theoretical basis for the IR extrapolation
was proposed in Ref.~\cite{Furnstahl:2012qg} (together with a model
for combined IR and UV extrapolations), and further developed in
Ref.~\cite{More:2013rma}.  

These papers demonstrate that oscillator
basis truncations (and more general basis truncations) effectively
impose a Dirichlet boundary condition (bc) at a properly identified
radius $L$ in position space. The radius $L$ is related to the
smallest eigenvalue $\kappa^2$ of the squared momentum operator
$\hat{p}^2$ in the finite basis, and $\kappa=\pi/L$.  For the
oscillator basis with highest excitation energy $N\hbar\Omega$, a very
accurate approximation is~\cite{More:2013rma}
\beq
\label{eq:L2_def}
  L = L_2\equiv\sqrt{2(N+3/2+2)}b \;,
\eeq
where $b=\sqrt{\hbar/(\mu\Omega)}$ is the oscillator length for a
particle with (reduced) mass $\mu$ and an oscillator frequency
$\Omega$.  The maximum excitation energy of the single-particle basis
is $N=2n+l$ in terms of the radial quantum number $n$ and the angular
momentum $l$. Note that $L_2$ differs slightly from the naive estimate
$L_0=\sqrt{2(N+3/2)}b$.  In localized bases that differ from the
harmonic oscillator, $L$ can be determined from a numerical
diagonalization of the operator $\hat{p}^2$.

\begin{figure}[bht!]
\centering
\includegraphics[width=0.95\columnwidth]{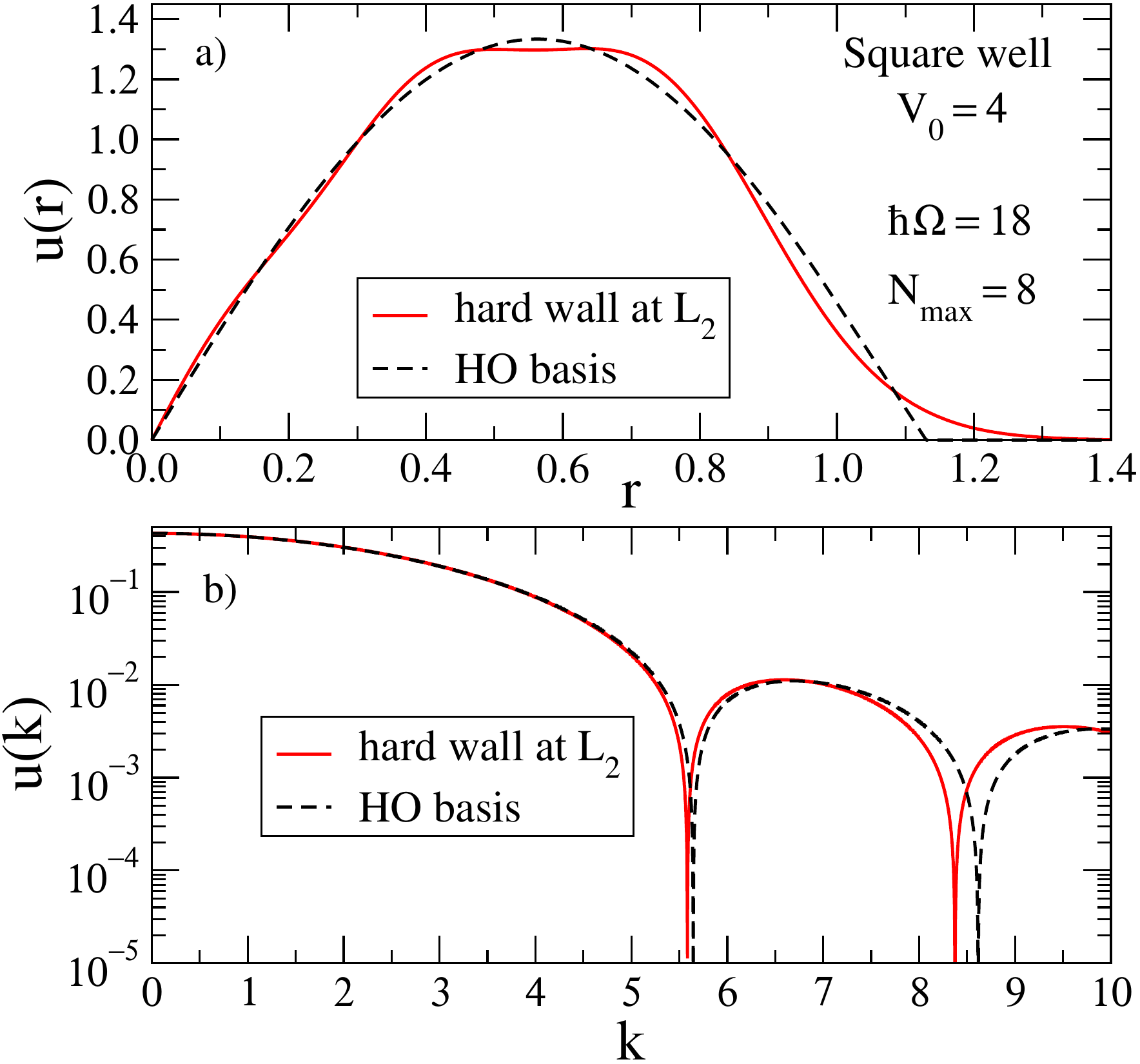}
\caption{(color online) Ground-state wave functions for a square well potential of depth
  $V_0=4$ (see Eq.~\eqref{eq:sq_well_def}; lengths are in units of $R$ and energies in units of
  $1/R^2$ with $\hbar^2/\mu = 1$) from solving the Schr\"odinger
  equation with a truncated harmonic oscillator basis with $\hbar\Omega = 18$
  and $N = 8$ (dashed) and with a
  Dirichlet boundary condition at $r=L_2$ given from Eq.~\eqref{eq:L2_def} (solid). 
  The coordinate-space radial wave functions in a) exhibit a difference at $r$
  near 1.5, but the Fourier-transformed wave functions in b) are in close
  agreement at low $k$, showing that the differences are high-momentum modes.
  }
\label{fig:HO_q_and_rspace_sqwell}
\end{figure}

The diagonalization of $\hat{p}^2$ shows that its low-lying spectrum in a finite
oscillator basis resembles that of a particle in a spherical
cavity of radius $L_2$~\cite{More:2013rma}. Therefore, the use of a Dirichlet boundary
condition at $L_2$ is a very convenient way to understand the long-wavelength
consequences of a finite basis. The difference between a Dirichlet bc and
the real asymptotic behavior of oscillator wave functions are high-momentum
modes, as can be shown by considering
Fourier transforms of the low-lying eigenstates
(an example is given in Fig.~\ref{fig:HO_q_and_rspace_sqwell}). 
Thus, this difference is
irrelevant for long-wavelength physics of bound states. We note that the use
of a Dirichlet bc is similar in spirit to the use of contact interactions
to describe the effect of unknown short-ranged forces on
long-wavelength probes.

A Dirichlet bc at $r=L$ allows one to derive formulas to extrapolate
bound-state energies and radii to infinite basis, and to predict
scattering phase shifts from the finite model space
results~\cite{More:2013rma}. In a simple view, the Dirichlet bc
introduces too much curvature into a bound-state wave
function, and the corresponding change in kinetic energy can be
derived accordingly. For applications and tests of the extrapolations
formulas we refer the reader to
Refs.~\cite{Soma:2012zd,Hergert:2012,Jurgenson:2013,Saaf:2013asa,Roth:2013fqa}.

We note that the IR extrapolation formulas~\cite{Furnstahl:2012qg} attain
for the oscillator basis (or any localized finite basis) what
L\"uscher's formula~\cite{Luscher:1985dn} achieves for the lattice.
The L\"uscher method has been extended to two-body bound
states with many recent developments 
(e.g., see Refs.~\cite{Lee:2010km,Zohreh:2011,Konig:2011nz,Pine:2013,Briceno:2013}). 
Here, the
oscillator basis has the advantage that two-body bound-state
extrapolations (e.g., see Ref.~\cite{More:2013rma}) are technically
not more complicated than for one-body systems.
In a discrete variable representation, IR and UV errors can also be
accessed conveniently~\cite{Bulgac:2013mz}.

The extrapolation formulas of Refs.~\cite{Furnstahl:2012qg} and
\cite{More:2013rma} were derived in a model-independent way based on
the so-called linear energy method~\cite{Djajaputra:2000aa}. In the
present work we reconsider the problem of a two-body system with
Dirichlet bc at $L$ to construct a consistent and systematic expansion
for the bound-state energy $E_L$ using a more general method based on
expanding the S-matrix about the bound-state pole in complex momentum.
This approach can be directly applied beyond $s$-waves and to coupled
channels, and manifests that $E_L$ depends only on observables.  We
extend the results in \cite{More:2013rma} to next-to-leading order
(NLO), correcting an inconsistent higher-order formula, and
demonstrate well-defined theoretical uncertainties for model problems
and a realistic deuteron calculation that uses a truncated oscillator basis
(with a range of oscillator parameters chosen
so that the UV contamination is negligible).

The plan of the paper is as follows.  In Sec.~\ref{sec:kL_S_matrix}
we use an analytic continuation of the S-matrix to complex momentum to
derive a transcendental equation for the $s$-wave binding momentum
$k_L$ for a Dirichlet bc at $L$.  We expand about the pole to derive
energy corrections up to NLO and validate the formulas using both
shallow and deep square wells as model test cases and calculations of
the deuteron that use a realistic interaction.  In
Sec.~\ref{sec:higher_l} we extend the results of
Ref.~\cite{More:2013rma} to identify the appropriate $L$ for orbital
angular momentum $l>0$ and generalize the energy extrapolation
formulas accordingly.  These extrapolations are tested in simple models, and with $l=2$
corrections to the deuteron results from Sec.~\ref{sec:kL_S_matrix}.
We show how the linear energy method can be used to reproduce the NLO
formula and introduce a new differential method in
Sec.~\ref{sec:derivative_method}.  
A derivation of correction formulas for the $s$-wave radius is
given in Sec.~\ref{sec:radii}.
In Sec.~\ref{sec:summary} we
summarize our results and discuss open questions on extensions to
other observables, $A>2$, and UV corrections.


\section{General $s$-wave equation for binding
  momentum}\label{sec:kL_S_matrix}

In this Section we derive an equation that determines the binding
momentum $k_L$ by relating the constraint of a Dirichlet bc at $r=L$
on the bound-state wave function to an analytic continuation of the
full-space S-matrix.  This relation allows us to expand momentum and
energy corrections order-by-order in terms of observables.  To
demonstrate this we use an effective range expansion and find the
corrections to NLO. This exercise also underscores the importance of
choosing a unitary form for the S-matrix to get correct higher-order
corrections.  Finally we test the analytical results obtained through
numerical studies of model potentials and a realistic deuteron.  

\subsection{Correction formulas to NLO}

The solution to the $s$-wave radial Schr\"odinger equation 
for the particular energy $E_L \equiv -k_L^2/2$ (with $\hbar^2/\mu =1$) 
for which the wave function vanishes at $r=L$ 
can be written in the asymptotic region (where the potential 
of range $R$ is negligible) as 
 \beq
  u_L(r) \overset{r \gg R}{\longrightarrow}  \left(e^{-k_L r} - e^{-2k_L L}e^{k_L r}\right)
   \;.
   \label{eq:uLasymp2}
\eeq
%
%
%
The relative coefficient of the two terms is uniquely fixed by the
boundary condition. (We note that the normalization is not relevant
here but will be considered below.) On the other hand, we can
analytically continue the asymptotic solution for positive energy to
complex momentum $ik_E$ in terms of the $s$-wave Jost function
$\Jost_0(k)$~\cite{taylor2006scattering}. This yields
\beq
   u_E(r)
       \overset{r \gg R}\longrightarrow
         \left(
           e^{-k_E r} - \frac{\Jost_0(ik_E)}{\Jost_0(-ik_E)} e^{k_E r}
         \right)
        \;,
  \label{eq:uLasymp3}
\eeq
with $E \equiv -k_E^2/2 < 0$.  For the particular energy $E = E_L$
where this has a zero at $r=L$ (for which $k_E = k_L$),
Eqs.~\eqref{eq:uLasymp2} and \eqref{eq:uLasymp3} must be the same wave
function, so
\beq
  e^{-2 k_L L} = \frac{\Jost_0(i k_L)}{\Jost_0(-i k_L)}
  \;.
  \label{eq:leads_to_basiceq}
\eeq
Moreover, this ratio of Jost functions gives the partial-wave S-matrix
for any $l$~\cite{taylor2006scattering}
\beq
   s_l(k) = \frac{\Jost_l(-k)}{\Jost_l(+k)}
   \;.
   \label{eq:s_l}
\eeq
Thus, the relation of the binding momentum to
the continuation of the $l=0$ S-matrix is
\beq
 e^{-2 k_L L} = [s_0(i k_L)]^{-1}
  \;.
  \label{eq:basiceq}
\eeq 

It remains to find an (approximate) expression or expansion for $s_0$
valid in this region of complex $k$, so that we can solve the above
transcendental equation for $k_L$ and thereby find $E_L$.

If the potential has no long-range part that introduces a singularity
in the complex $k$ plane nearer to the origin than the bound-state
pole (which is the case, for example, for the deuteron when we assume that
the longest-ranged interaction is from pion exchange), then the
continuation of the positive-energy partial-wave S-matrix (i.e., the
phase shifts) to the pole should be unique.  Because $|k_L| <
|\kinf|$, $s_0(i k_L)$ and therefore $k_L$ and the energy shift $E_L$
should be determined solely by observables.

The leading term in an expansion of $k_L - \kinf$ using
Eq.~\eqref{eq:basiceq} comes from the bound-state pole, at which $s_0$
behaves like~\cite{newton2002scattering}
\beq
  s_0(k) \approx \frac{-i\ANC^2}{k-i\kinf}
  \;.
  \label{eq:purepole}
\eeq
Here $\ANC$ is
the asymptotic normalization coefficient (ANC).  The ANC is defined by the large-$r$ behavior of the
\emph{normalized} bound-state wave function
\beq
  u_{\rm norm}(r)\overset{r\gg R}\longrightarrow \ANC e^{-\kinf r}
  \;.
  \label{eq:definition_of_ANC}
\eeq
Substituting Eq.~\eqref{eq:purepole} into Eq.~\eqref{eq:basiceq} yields
\beq
  k_L - \kinf \approx -\ANC^2 e^{-2k_L L} \approx -\ANC^2 e^{-2\kinf L}
  \;.
  \label{eq:kLatLO}
\eeq
This is the leading-order (LO) result for $k_L$ obtained earlier in 
Ref.~\cite{More:2013rma}.

Iterations of the intermediate equation  in \eqref{eq:kLatLO} as well as 
the results from Ref.~\cite{More:2013rma} motivate the NLO parameterization of $k_L$ as
\beq
  k_L = \kinf + A e^{-2 \kinf L} + (B L + C) e^{-4 \kinf L}  + \mathcal{O}(e^{-6 \kinf L})
  \;,
  \label{eq:kLexpansion}
\eeq
with $A = -\ANC^2$.  In general we can substitute this expansion into
Eq.~\eqref{eq:basiceq} using an parametrized form of the S-matrix,
then expand in powers of $e^{-2 \kinf L}$ and equate $e^{-2 \kinf L}$,
$L e^{-4 \kinf L}$, and $e^{-4 \kinf L}$ terms on both sides of the
equation.  However, while both $A$ and $B$ are uniquely determined by
the pole in $s_0(k)$ at $k=i\kinf$, $C$ is only determined unambiguously if
$s_0(k)$ is consistently parameterized away from the pole.  For
example, the two parametrizations
\beq
\label{eq:S0_1}
    s_0(i k_L) \approx \frac{\kinf^2 - k_L^2 + 2 k_L \ANC^2}{\kinf^2 - k_L^2}
\eeq
and
\beq
  s_0(k) \approx \frac{-\ANC^2}{2\kinf}\,\frac{k+i\kinf}{k- i\kinf}
  \label{eq:s0_Newton}
\eeq
yield different results for $C$. The first
parametrization~(\ref{eq:S0_1}) is based on a particular form for the
partial-wave scattering amplitude near the
pole~\cite{taylor2006scattering}, and was employed in
Ref.~\cite{More:2013rma}.  The second paramerization
(\ref{eq:s0_Newton}) correctly incorporates that the S-matrix also has
a zero at $-i\kinf$~\cite{newton2002scattering}.  In neither case,
however, do we have a sufficiently general parametrization that allows
us to unambiguously determine $C$.

For the complete NLO energy correction, we start from the
general expression for the S-matrix
\beq
  s_0(k) = \frac{k\,\cot\delta_0(k) + ik}{k\,\cot\delta_0(k) -ik}
  \;,
  \label{eq:s0_exact}
\eeq
and use an effective range expansion to substitute for $k\,\cot
\delta_0(k)$.  In particular, we use an expansion
around the bound-state pole rather than about zero energy, 
namely~\cite{wu2011scattering,Phillips:1999hh},
 \beq
  k\,\cot\delta_0(k) = -\kinf + \frac12 \rho_d(k^2 + \kinf^2)
     + w_2(k^2+\kinf^2)^2 +  \cdots
     \;.
     \label{eq:eff_range_kinf}
\eeq 
To match the residue at the S-matrix pole as in \eqref{eq:purepole},
we identify
\beq
 \rho_d = \frac{1}{\kinf} - \frac{2}{\ANC^2}
  \;.
  \label{eq:rhoD_gamma_rel}
\eeq
Now we substitute \eqref{eq:eff_range_kinf} into \eqref{eq:s0_exact}
and use Eq.~\eqref{eq:kLexpansion} to expand both sides of
Eq.~\eqref{eq:basiceq}, equating terms with equal powers of
$e^{-2\kinf L}$ and $L$. The resulting expansion for the binding
momentum to NLO is
\begin{align}
    [k_L]_{\rm NLO} &=  \kinf - \ANC^2 e^{-2\kinf L}  - 2  L \ANC^4  e^{-4\kinf L} 
    \nonumber  \\
      &\null - \ANC^2 \left(1 - \frac{\ANC^2}{2\kinf} - \frac{\ANC^4}{4\kinf^2} + 2 \kinf w_2 \ANC^4 \right) e^{-4\kinf L}
      \;.
    \label{eq:complete_k_correction_NLO}
\end{align}
Using $\Delta E_L \equiv E_L - \Einf = \kinf^2/2 - k_L^2/2$, 
the correction for the energy due to finite $L$ is
\begin{align}
  [\Delta E_L]_{\rm NLO} &=  \kinf \ANC^2 e^{-2\kinf L}
  + 2\kinf L\ANC^4 e^{-4\kinf L} 
    \nonumber \\
  & \null\quad +
    \kinf\ANC^2 \Bigl( 1-\frac{\ANC^2}{\kinf}-\frac{\ANC^4}{4\kinf^2} 
    \nonumber \\
  & \null\quad +
     2\kinf w_2 \ANC^4  \Bigr)  e^{-4\kinf L}
      \;.  
      \label{eq:complete_E_correction_NLO}
\end{align}
In what follows we use LO to refer to the first term in this expansion and L-NLO to
refer to the first two terms (the second term should dominate the full
NLO expression when $\kinf L$ is large).  We also note that
higher-order terms in Eq.~\eqref{eq:eff_range_kinf} (e.g., terms
proportional to $(k^2+\kinf^2)^3$ and higher powers) do not affect the
binding momentum or energy predictions
Eqs.~\eqref{eq:complete_k_correction_NLO} and
\eqref{eq:complete_E_correction_NLO} at NLO.

As a special case, let us consider the zero-range limit of a
potential. In this case $\rho_d = w_2 = 0$, $\ANC^2 = 2\kinf$, and 
\beq
   [s_0(ik_L)]^{-1} = \frac{\kinf-k_L}{\kinf+k_L}\;.
\eeq
The expansion for $k_L$ in a form similar to Eq.~\eqref{eq:kLexpansion}
can be extended to arbitrary order using Eq.~\eqref{eq:basiceq}.

We note finally that the leading corrections beyond NLO scale as $L^2
e^{-6\kinf L}$. While we do not pursue a derivation of such high-order
corrections here, the knowledge of the leading form is useful in some of
the error analysis we present below.

\subsection{Numerical tests}\label{subsec:test}

In this Subsection we test the expansion for $\Delta E_L$ for an
analytically solvable model and also consider the deuteron based on
realistic nucleon-nucleon intercations. For the square-well potential
\beq
  V_{\rm sw} (r) = -V_0\, \theta (R-r)\;,
  \label{eq:sq_well_def}
\eeq
the parameters in Eq.~\eqref{eq:complete_E_correction_NLO}
can be calculated exactly.  The $s$-wave scattering phase shift for
the square well is
\beq 
 \delta_0 (k) =
  \tan^{-1}\left[\sqrt{{k^2 \over k^2 + \eta^2}} \tan(\sqrt{k^2+\eta^2}
  R)\right] - k R\;, 
  \label{eq:delta0}
\eeq 
with $\eta = \sqrt{2 V_0}$.  Analytically
continuing the effective range expansion by taking $k \rightarrow i k_L$ 
in Eqs.~\eqref{eq:eff_range_kinf} and \eqref{eq:delta0}, we obtain
\begin{widetext}
\beq
   {i k_L \sqrt{\eta^2 - k_L^2} -k_L^2 \tan( \sqrt{\eta^2 - k_L^2} R) 
  \tan(i k_L R) \over i k_L \tan(\sqrt{\eta^2 - k_L^2} R) - \sqrt{\eta^2 - k_L^2} \tan(i k_L R) } 
  = -\kinf  
+ \frac12{\rho_d} (\kinf^2-k_L^2) + w_2 (\kinf^2 - k_L^2)^2 + \mathcal{O}\left((\kinf^2-k_L^2)^3\right)
\;. 
\label{eq:get_w2_sq_well}
\eeq
\end{widetext}
The branch for the square-root is specified by the requirement that
$\tan \delta (i\kinf) = -i$.  
To get $\rho_d$ ($w_2$) we differentiate once (twice) each side with
respect to $k_L$ and then set $k_L = \kinf$.  The $\rho_d$ obtained in
this way is consistent with Eq.~\eqref{eq:rhoD_gamma_rel} when $\ANC$
is obtained by the large $r$ behavior of the bound-state wave function
as defined in Eq.~\eqref{eq:definition_of_ANC}.

In addition, the square well with a Dirichlet bc at $L>R$ can be solved
systematically for the binding momentum.  The matching condition
yields 
\beq \kappa_L \cot \kappa_L R = - k_L \frac{1 + e^{-2k_L
    (L-R)}}{1-e^{-2k_L (L-R)}} \;,\label{eq:sq_well_matchingL} 
\eeq
where $k_L = \sqrt{2 |E_L|}$ and $\kappa_L = \sqrt{\eta^2 - k_L^2}$.
We expand both sides of Eq.~\eqref{eq:sq_well_matchingL} in powers of
\beq 
  \Delta k \equiv k_L - \kinf\;.
  \label{eq:Delta_k_def}
\eeq
We write the left-hand side of Eq.~\eqref{eq:sq_well_matchingL} as
\beq
\label{eq:LHS_matchingL_expansion}
\kappa_L \cot \kappa_L R = \kappainf \cot \kappainf R + \mathcal{A} (\Delta k) 
   + \mathcal{B} (\Delta k)^2+\cdots \;,
\eeq
and obtain the coefficients $\mathcal{A}$, $\mathcal{B}$ by Taylor expanding $\kappa_L
\cot (\kappa_L R)$ around $\kinf$. We write $\Delta k$ as 
\beq
\label{eq:Delta_k_def2} \Delta k = k_{(1)} + k_{(2)} + \cdots \;.
\eeq
Here $k_{(1)} \sim e^{-2 \kinf L}$ is the LO correction, $k_{(2)}
\sim e^{-4\kinf L}$ is the NLO correction and so on, and we truncate the
expressions consistently to obtain the energy correction for the
square well to the desired order.  The results of the general S-matrix
and square-well-only Taylor expansion methods of calculating energy
corrections are found to match explicitly at LO, \LNLO, and NLO.

\begin{figure}[t!]
\centering
\includegraphics[width=0.95\columnwidth]{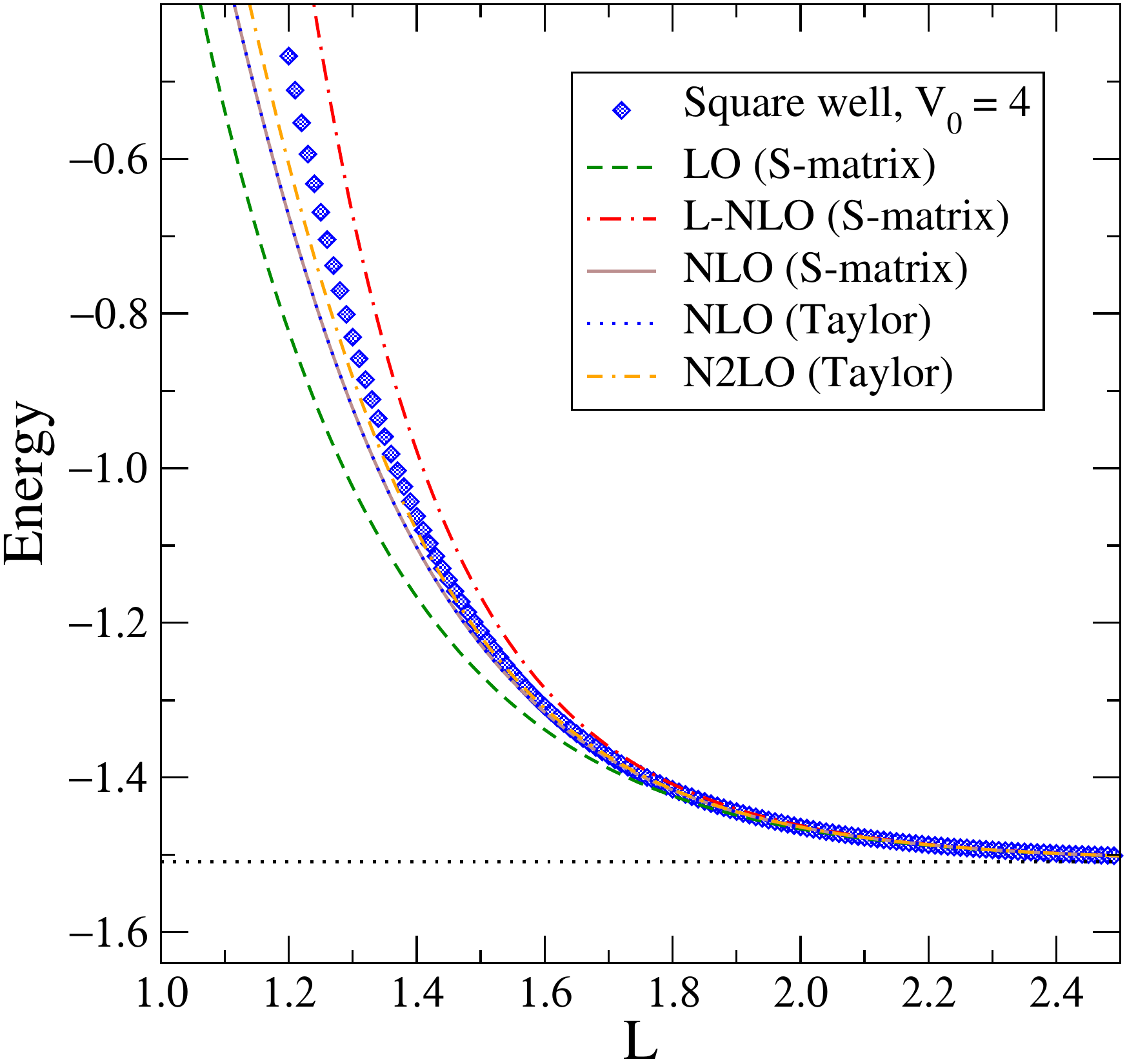}
\caption{(color online) Bound-state energy for a square well of depth
  $V_0=4$ (lengths are in units of $R$ and energies in units of
  $1/R^2$ with $\hbar^2/\mu = 1$) from solving the Schr\"odinger
  equation with a Dirichlet boundary condition at $r=L$.  The diamonds
  are exact results for each $L$ while the horizontal dotted line is
  the energy for $L\rightarrow\infty$, $\Einf = -1.5088$.  The dashed,
  dot-dashed and solid lines are predictions for the energy using the
  systematic correction formula
  Eq.~\eqref{eq:complete_E_correction_NLO} at LO (first term only),
  \LNLO\ (first two terms), and full NLO (all terms), respectively.
  The dotted curve on top of the solid line and the dot-double-dashed
  lines are respectively the NLO and N2LO predictions for the square
  well from the Taylor expansion method described in
  Sec.~\ref{subsec:test}.}
\label{fig:sqwell_curvesV4}
\end{figure}

\begin{figure}[t!]
\centering
\includegraphics[width=0.95\columnwidth]{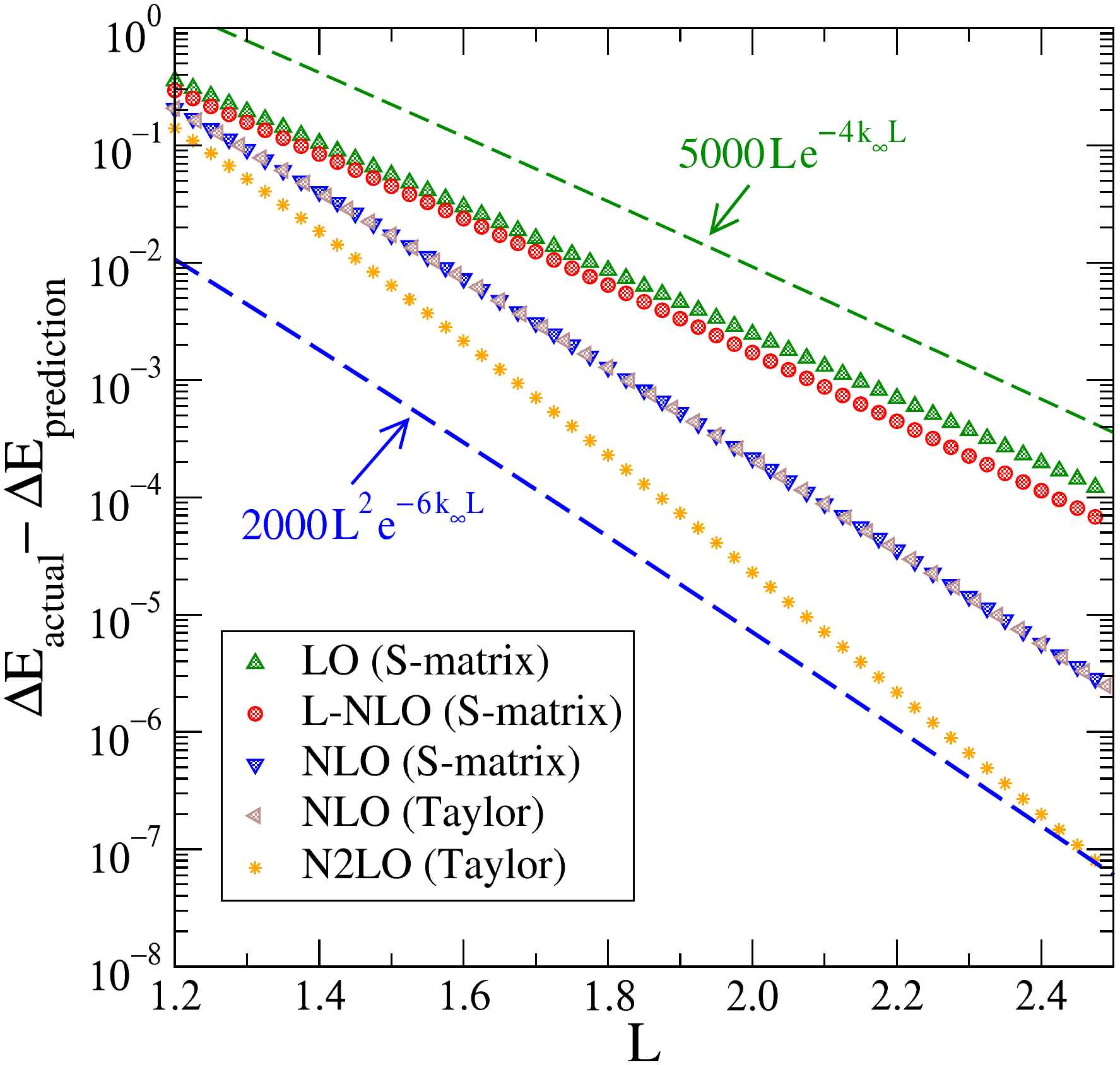}
\caption{(color online) Error plots of the energy correction at each
  $L$ for the square well of Fig.~\ref{fig:sqwell_curvesV4} ($V_0 =
  4$) predicted at different orders by
  Eq.~\eqref{eq:complete_E_correction_NLO} and by the Taylor expansion
  method in Sec.~\ref{subsec:test}, each compared to the exact energy.
  Lines proportional to $L e^{-4\kinf L}$ (dashes) and $L^2 e^{-6\kinf
    L}$ (with arbitrary normalization) are plotted for comparison to
  anticipated error slopes. }
\label{fig:sq_well_error_log_plotsV4}
\end{figure}

Figure~\ref{fig:sqwell_curvesV4} compares the energy corrections
for the general S-matrix method at LO and NLO for a representative
square-well potential with one bound state to the exact energies.  The
Taylor expansion results for the square well at NLO and N2LO (which is
proportional to $e^{-6\kinf L}$) are also plotted.  We note that the
predictions are systematically improved as higher-order terms are
included and that keeping terms only up to \LNLO~overestimates the
energy correction.  Also as seen in Fig.~\ref{fig:sqwell_curvesV4},
the full NLO energy correction predicted by
Eq.~\eqref{eq:complete_E_correction_NLO}, with $w_2$ determined by
Eq.~\eqref{eq:get_w2_sq_well}, matches the `exact' NLO result obtained
by Taylor expansion.  This confirms that
Eq.~\eqref{eq:complete_E_correction_NLO} is indeed the complete energy
correction at NLO.

To see if the errors decrease with the implied systematics, we plot
the difference of actual energy corrections and the energy corrections
predicted at different orders on a log-linear scale in
Fig.~\ref{fig:sq_well_error_log_plotsV4}. We observe that the errors
successively decrease at each fixed $L$ as we go from LO to NLO to
N2LO. The up triangles in Fig.~\ref{fig:sq_well_error_log_plotsV4} are
$\Delta E_{\rm actual} - \Delta E_{\rm LO}$.  From
Eq.~\eqref{eq:complete_E_correction_NLO} the dominant omitted
correction in $\Delta E_{\rm LO}$ is proportional to $L e^{-4 \kinf
  L}$.  As seen in Fig.~\ref{fig:sq_well_error_log_plotsV4}, the slope
of $\Delta E_{\rm actual} - \Delta E_{\rm LO}$ is roughly $L
e^{-4 \kinf L}$, as expected.  We also note that $\Delta E_{\rm \LNLO}$
is only a marginal improvement over $\Delta E_{\rm LO}$ for the
plotted range of $L$ and that
$\Delta E_{\rm actual} - \Delta E_{\rm NLO}$ has the expected slope of
$L^2 e^{-6 \kinf L}$. We again see a perfect agreement between the
results obtained from the S-matrix method
\eqref{eq:complete_E_correction_NLO} and those obtained from the
Taylor expansion of Eq.~\eqref{eq:sq_well_matchingL}.  We have also
studied deeper square wells with more than one bound state
and verified that the S-matrix approach applies to all the bound states.

\begin{figure}[tbh!]
\centering
\includegraphics[width=0.95\columnwidth]{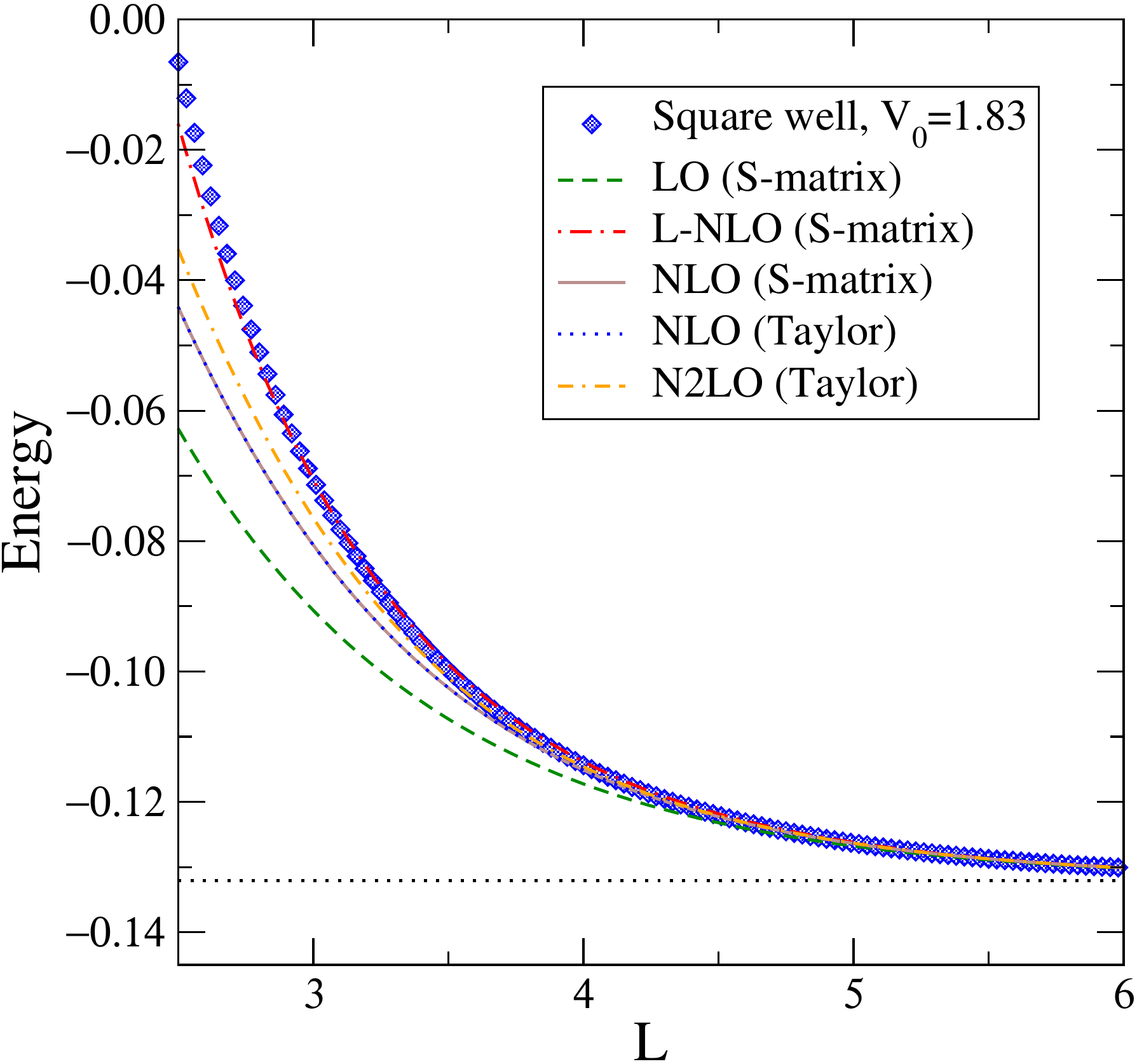}
\caption{(color online) Bound-state energy for a square well of depth
  $V_0=1.83$ (units with $R=1$), which simulates a deuteron, from
  solving the Schr\"odinger equation with a Dirichlet boundary
  condition at $r=L$.  The horizontal dotted line is the exact energy,
  $\Einf = -0.1321$ and the other curves are as the same as in
  Fig.~\ref{fig:sqwell_curvesV4}.}
\label{fig:sqwell_curves}
\end{figure}

\begin{figure}[tbh!]
\centering
\includegraphics[width=0.95\columnwidth]{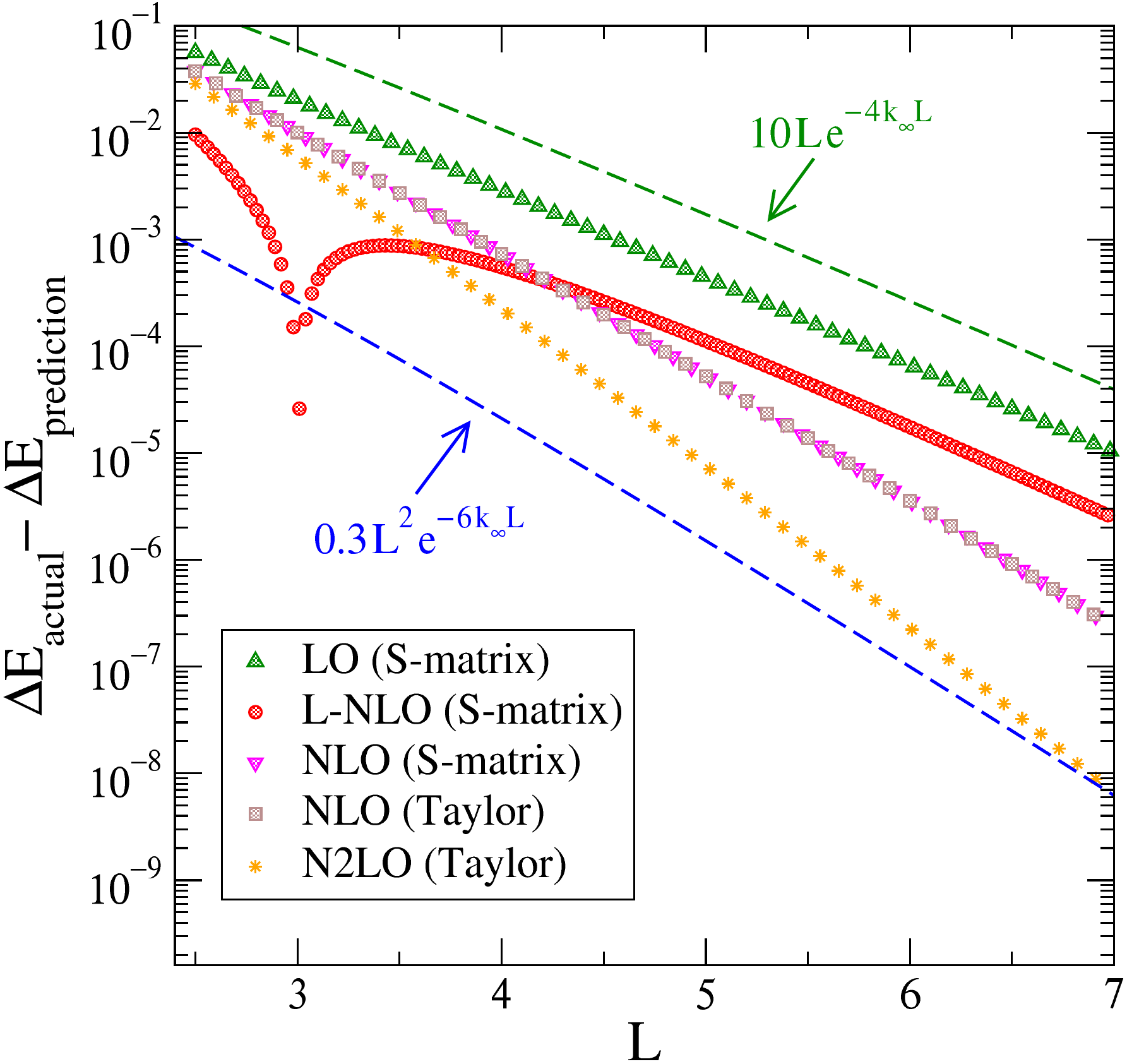}
\caption{(color online) Comparison of the actual energy correction due
  to truncation to the energy correction predicted to different orders
  by Eq.~\eqref{eq:complete_E_correction_NLO} for a square well
  (Eq.~\eqref{eq:sq_well_def}) with $V_0 = 1.83$ and $R=1$.}
\label{fig:sq_well_error_log_plots}
\end{figure}

In Figs.~\ref{fig:sqwell_curves} and
\ref{fig:sq_well_error_log_plots}, the same analysis is done but now
with the depth of the square well adjusted so that the exact binding
energy is the same as the deuteron binding energy scaled to the units
$\hbar=1$, $\mu=1$ and $R=1$. An important difference in this case
compared to the deeper square well is that the \LNLO~prediction gives
a very close estimate for the truncated energies at smaller $L$
values.  
However, the small errors in this region should not be over-emphasized
because they are not systematic.
As seen in
Fig.~\ref{fig:sq_well_error_log_plots},
$\Delta E-\Delta E_{\rm \LNLO}$ is the dominant NLO correction  at large $L$
but still has about the same slope as $\Delta E -
\Delta E_{\rm LO}$, reflecting the $L$-dependence of the 
remainder of the NLO correction.  
Only when the full NLO correction is included does the slope
go to $L^2 e^{-6 \kinf L}$.

\begin{figure}[tbh!]
\centering
\includegraphics[width=0.95\columnwidth]{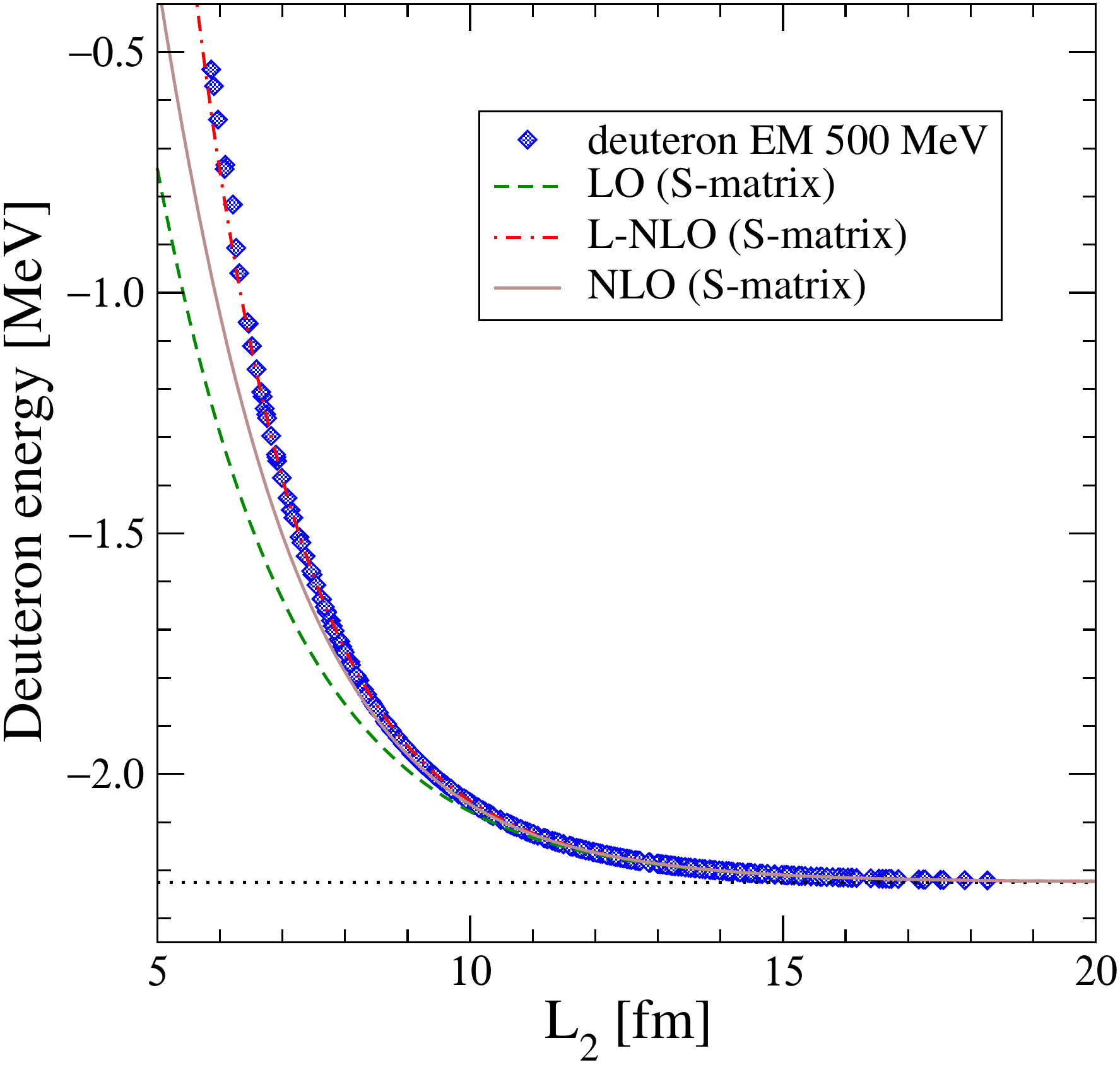}
\caption{(color online) Deuteron energy versus $L_2$ (see
  Eq.~\eqref{eq:L2_def}) for the chiral N$^3$LO (500\,MeV) potential
  of Ref.~\cite{Entem:2003ft}.  To eliminate the UV contamination we
  only plot results for $\hw > 49$~MeV.  The dashed, dot-dashed and solid
  lines are respectively the LO (first term in
  Eq.~\eqref{eq:complete_E_correction_NLO}), \LNLO~(first two terms in
  Eq.~\eqref{eq:complete_E_correction_NLO}) and the full NLO (all the
  terms in Eq.~\eqref{eq:complete_E_correction_NLO}) predictions for
  the energy correction.  The horizontal dotted line is the deuteron
  energy.}
\label{fig:deuteron_curves}
\end{figure}

\begin{figure}[tbh!]
\centering
\includegraphics[width=0.95\columnwidth]{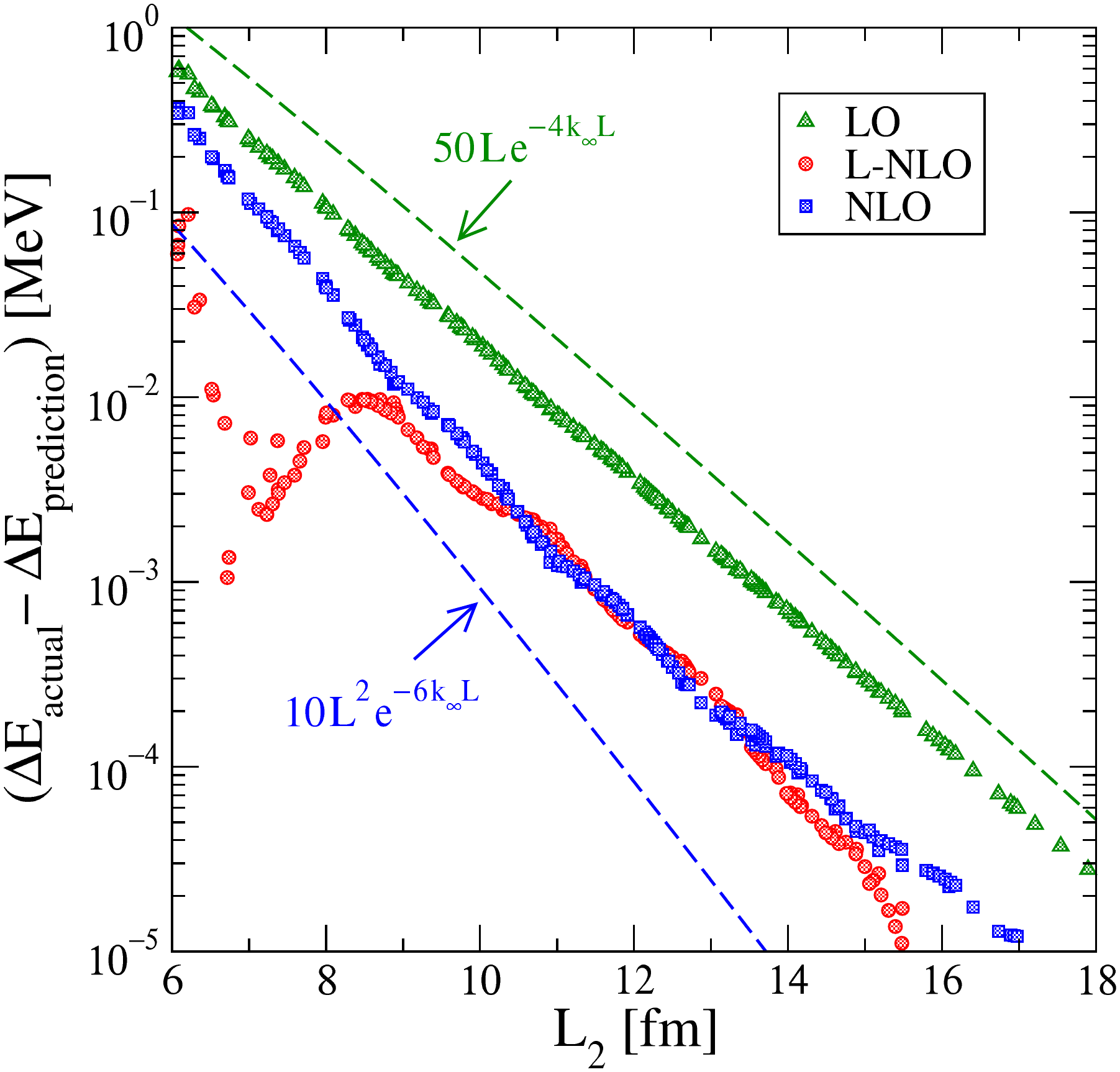}
\caption{(color online) Comparison of the actual energy correction due
  to HO basis truncation ($\hw$ restricted to be greater than $49$~MeV to
  eliminate UV contamination) for the deuteron to the energy
  correction predicted to different orders from
  Eq.~\eqref{eq:complete_E_correction_NLO}. For the parameter $w_2$ in
  Eq.~\eqref{eq:complete_E_correction_NLO} we use the value reported
  in \cite{Phillips:1999hh}.}
\label{fig:deuteron_error_log_plots}
\end{figure}

Figures~\ref{fig:deuteron_curves} and
\ref{fig:deuteron_error_log_plots} show analogous results for the
deuteron calculated with the chiral EFT potential of
Ref.~\cite{Entem:2003ft}.  We use the HO basis and predict the ($l=0$) energy
correction from Eq.~\eqref{eq:complete_E_correction_NLO} assuming a
Dirichlet bc at $L_2$ given by Eq.~\eqref{eq:L2_def}.  We only include
energies for which $\hw > 49$~MeV, which is sufficient to render UV
corrections negligible.  For the parameter $w_2$ in
Eq.~\eqref{eq:complete_E_correction_NLO} we use $w_2
= 0.389$ as reported in \cite{Phillips:1999hh}.  We also note that the
$\rho_d$ value reported in \cite{Phillips:1999hh} satisfies
Eq.~\eqref{eq:rhoD_gamma_rel}, where $\ANC$ now is the $s$-wave ANC.
The $y$-axis minimum is dictated by the limited precision of
the ANC and $w_2$ values.
We notice again that the close agreement of the \LNLO~prediction to
the deuteron data is not systematic while the full corrections to the
LO and NLO predictions have the anticipated slopes except at large
$L_2$.  In the next Section we extend our formulas to $l>0$, which
enables us to include contributions from the $d$-wave at LO.  This
becomes noticable on the error plot for large $L_2$ (see
Fig.~\ref{fig:deuteron_error_plot_l2}).


\section{Extension to nonzero orbital angular momentum}\label{sec:higher_l}

The deuteron ground state is a mixture of an $s$ and a $d$ state, and
the $s$ and $d$ asymptotic normalization coefficients (as well as the
$d$-to-$s$ state ratio of about 2.5\%) are observables.  The
extrapolation formulas in the previous Section were derived for $s$
states, and it is of interest to extend these to nonzero angular
momenta $l$. We do so in two steps. First, we show that $L_2$ is also
the relevant effective hard-wall radius for oscillator wave functions
with nonzero angular momenta. Second, we derive the energy correction
for nonzero angular momenta.

\subsection{$L$ for nonzero angular momenta}

For the derivation of the relevant IR length scale at $l>0$ we closely
follow Ref.~\cite{More:2013rma}. We compute the smallest eigenvalue
$\kappa^2$ of the squared momentum operator $\hat{p}^2$ in a finite
oscillator basis and identify $\kappa =x_l/L$ (with $x_l$ being the
smallest positive zero of the spherical Bessel function $j_l$). This
identification, and the form of the corresponding eigenfunctions are,
of course, guided by the Dirichlet bc at $r=L$. Throughout this
Section, we set the oscillator length $b=1$.  
Because this is the only length scale here, the results are general 
and can be extended to any $b$ with a simple rescaling.
The normalized radial
oscillator wave function of energy
\beq
  E=2n+l+3/2
\eeq
is   $\psi_{nl}(r)=u_{nl}(r)/r$ with
\beq
u_{nl}(r)=\sqrt{2n!\over\Gamma(n+l+3/2)} r^{l+1} e^{-r^2/2} L_{n}^{l+1/2}(r^2) \;.
\eeq
Here, $L_n^{l+1/2}$ denotes the generalized Laguerre polynomial. 

In this basis, the operator $\hat{p}^2$ of the momentum squared is
tridiagonal with matrix elements
\begin{align}
\langle u_{ml}|\hat{p}^2|u_{nl}\rangle &= (2n+l+3/2)\delta_m^n \nonumber\\
   & \qquad \null +\sqrt{n+1} 
   \sqrt{n+l+3/2}\,\delta_m^{n+1} \nonumber \\
  & \qquad \null +\sqrt{n}\sqrt{n+l+1/2}\,\delta_m^{n-1} \;.
\end{align}
For the eigenfunction of $\hat{p}^2$ with smallest eigenvalue
$\kappa^2$ at angular momentum $l$, we make the ansatz $\psi_{\kappa
  l}(r)/r$ with
\bea
\label{eigen}
\psi_{\kappa l}(r) =\left\{\begin{array}{cc}
\kappa r j_l(\kappa r) \;, & 0\le \kappa r \le x_l \;, \\
0 \;, & \kappa r > x_l \;.\end{array}\right.  
\eea
Here, $j_l$ is the regular spherical Bessel function and $x_l$ is its
smallest positive zero. Clearly, these eigenfunctions are those of a
particle in a spherical cavity with a Dirichlet bc at $x_l/\kappa$. In
an infinite basis, the wave function $\psi_{\kappa l}(r)/r$ is an
eigenfunction of $\hat{p}^2$ for any non-negative value of $\kappa$.
In a finite oscillator basis, only discrete momenta $\kappa$ are
allowed. For their computation we expand the eigenfunction as
\beq
\psi_{\kappa l}(r) = \sum_{m=0}^{n} c_m(\kappa) u_{ml}(r) \;,
\eeq
where we supress the dependence of the admixture coefficients
$c_m(\kappa)$ on $l$, which is kept fixed throughout this derivation.

The last row of the matrix eigenvalue problem for $\hat{p}^2$ is
\beq
\label{quantize}
(2n+l+3/2 -\kappa^2)c_n(\kappa) = -\sqrt{n}\sqrt{n+l+1/2}\, c_{n-1} \;, 
\eeq 
and this becomes the quantization condition for $\kappa$. The direct computation of
the coefficients $c_n(\kappa)$ seems difficult. Instead, we make a
Fourier-Bessel expansion
\beq
\label{fourierbessel}
\psi_{\kappa l}(r)= \sqrt{2\over \pi}\int\limits_0^\infty\! dk\, 
   \tilde{\psi}_{\kappa l}(k)\, kr j_l(kr) \;, 
\eeq
and use 
\beq
kr j_l(kr) =\sqrt{\pi\over 2}\sum_{n=0}^\infty (-1)^n u_{nl}(k) u_{nl}(r) \;.
\eeq
Thus, 
\beq
\psi_{\kappa l}(r)= \sum_{n=0}^\infty (-1)^n u_{nl}(r)
\int\limits_0^\infty\! dk\, \tilde{\psi}_{\kappa l}(k) u_{nl}(k) \;, 
\eeq
and the admixture coefficients are therefore
\beq
c_n(\kappa)= (-1)^n \int\limits_0^\infty dk\, \tilde{\psi}_{\kappa l}(k) u_{nl}(k) \;.
\eeq

So far, our formal manipulations have been exact. We now employ an
asymptotic approximation of the generalized Laguerre polynomials
(which enters the $u_{nl}(k)$) in terms of Bessel functions, valid for
$n\gg 1$, see Eq.~(15) of Ref.~\cite{Deano2013}. This yields
\begin{align}
u_{nl}(k) &\approx {2^{1-n}\over \pi^{1/4}} \sqrt{(2n+2l+1)!\over (n+l)! n!} 
   \,(4n+2l+3)^{-{l+1\over 2}}
  \nonumber \\
  & \qquad \null\times \sqrt{4n+2l+3} k\, j_l(\sqrt{4n+2l+3}k) \;, 
\end{align}
and 
\begin{align}
c_n(\kappa) &\approx C_{nl} \sqrt{2\over \pi}\int\limits_0^\infty dk \, \tilde{\psi}_{\kappa l}(k) 
\sqrt{4n+2l+3} k \nonumber \\
  & \qquad\qquad\qquad \null \times j_l(\sqrt{4n+2l+3}k) \;. 
\end{align}
Here, $C_{nl}$ is a constant that does not depend on $\kappa$. The key
point is that the asymptotic expansion in terms of Bessel functions
allows us now to employ the definition~\eqref{fourierbessel} to 
evaluate the integral
\begin{align}
 \sqrt{2\over \pi}\int\limits_0^\infty dk\, & \tilde{\psi}_{\kappa l}(k) 
\sqrt{4n+2l+3} k\, j_l(\sqrt{4n+2l+3}k) \nonumber \\
&=  \psi_{\kappa l}(\sqrt{4n+2l+3}) \nonumber \\
&= \sqrt{4n+2l+3}\, \kappa\, j_l(\sqrt{4n+2l+3}\kappa) \;.
\end{align}
Putting it all together, we find
\begin{align}
c_n(\kappa) &= {2^{1/2-n}(-1)^n \pi^{1/4} \over (4n+2l+3)^{l/2}}
\sqrt{(2n+2l+1)!\over (n+l)!n!} \nonumber \\
 & \qquad\null\times
\kappa\, j_l(\sqrt{4n+2l+3}\kappa) \;.
\end{align}
We insert this expression for $c_n(\kappa)$ into the quantization
condition~\eqref{quantize} and make the ansatz
\beq
\kappa ={x_l\over\sqrt{4n+2l+3+2\Delta}} \ . 
\eeq
Assuming the limit $n\gg 1$ and $n\gg l$ in the quantization condition
then yields
\beq
\Delta=2 \ .
\eeq
Thus, $\Delta$ does not depend on $l$ in this limit, and the result
is consistent with the $l=0$ result of Ref.~\cite{More:2013rma}. In
other words, the extent of the position space in finite oscillator
basis with maximum radial quantum number $n$ and angular momentum $l$ is
\bea
\label{L2}
  L_2 &=& \sqrt{2(2n+l+3/2+2)}b \nonumber\\
&=& \sqrt{2(N+3/2+2)}b \;,
\eea
in accord with Eq.~\eqref{eq:L2_def}.    

Table~\ref{tab:lowest_kappa} shows numerical comparisons for $l=0,1,2$
and a range of $n$ of the exact minimum momentum $\kappa$ and the
estimate $x_l/L_2$ (with $x_0=\pi$, $x_1\approx 4.49341$, $x_2\approx
5.76346$).  The estimates are accurate approximations of the exact
results even for small $N=2n+l$, but the accuracy decreases somewhat
with increasing orbital angular momentum. In some practical
calculations it might thus be of advantage to directly employ the
numerical results for $L_2$ instead of the approximate analytical
expression~(\ref{L2}). 

\begin{table}[tbh]
\caption{Comparison of the exact lowest momentum $\kappa$ with the analytical 
estimate $x_l/L_2$ for $l=0, 1, 2$ and $0 \leq n \leq 10$.}
\label{tab:lowest_kappa}
\begin{tabular}{c|c|c|c||c|c|c|c||c|c|c|c}
$l$& $n$ & $\kappa$  &${x_l/ L_2}$ & 
$l$& $n$ & $\kappa$  &${x_l/ L_2}$ &
$l$& $n$ & $\kappa$  &${x_l/ L_2}$ \\
\hline
0 &  0  & 1.2247  & 1.1874   &
1 &  0  & 1.5811  & 1.4978   &
2 &  0  & 1.8708  & 1.7378   \\
0 &  1  & 0.9586  & 0.9472   &
1 &  1  & 1.2764  & 1.2463   &
2 &  1  & 1.5423  & 1.4881   \\
0 &  2  & 0.8163  & 0.8112   &
1 &  2  & 1.1047  & 1.0898   &
2 &  2  & 1.3509  & 1.3222   \\
0  &  3  &  0.7236  &  0.7207  &
1  &  3  &  0.9892  &  0.9805  &
2 &    3  &    1.2191  &     1.2018    \\
0   &  4   &   0.6568   &    0.6551    &
1 &    4 &     0.9042   &    0.8987  &
2 &    4  &    1.1207  &     1.1092   \\
0  &   5   &   0.6058   &    0.6046    &
1 &    5 &     0.8382   &    0.8344  &
2 &    5   &   1.0432   &    1.0352    \\
0  &   6   &   0.5651   &    0.5642    &
1 &    6 &     0.7850   &    0.7822  &
2  &   6  &    0.9801  &     0.9742  \\
0   &  7 &     0.5316    &   0.5310    &
1  &   7 &     0.7408   &    0.7387  &
2 &    7  &    0.9274  &     0.9229   \\
0  &   8    &  0.5035    &   0.5031    &
1  &   8 &     0.7033   &    0.7018  &
2 &    8  &    0.8824  &     0.8789   \\
0  &  9  &    0.4795   &    0.4791     &
1 &    9 &     0.6711   &    0.6698  &
2  &   9   &   0.8435  &     0.8407    \\
0  &   10 &   0.4585  &     0.4582    &
1 &    10 &   0.6429  &     0.6419   &
2  &   10   & 0.8093 &      0.8070  \\
\end{tabular}
\end{table}

\subsection{Energy correction for finite angular momentum}

Let us extend our $l=0$ result for $[\Delta E]_{\rm LO}$ to $l>0$ following the
method in Sec.~\ref{sec:kL_S_matrix}.
For orbital angular momentum $l$, the asymptotic wave function is
\beq
   u_L(r) \overset{r \gg R}{\longrightarrow}  k_L r\Bigl(h_l^{(1)}(ik_L r) - 
{h_l^{(1)}(ik_L L)\over h_l^{(1)}(-ik_L L)} 
h_l^{(1)}(-ik_Lr)\Bigr)
   \;.
   \label{eq:uLasympl}
\eeq
Here, $h_l^{(1)}$ denotes the spherical Hankel function of the first
kind (or the spherical Bessel function of the third
kind)~\cite{abramowitz1964}. By definition $u_L(L)=0$.

\begin{figure}[h!]
\centering
\includegraphics[width=0.95\columnwidth]{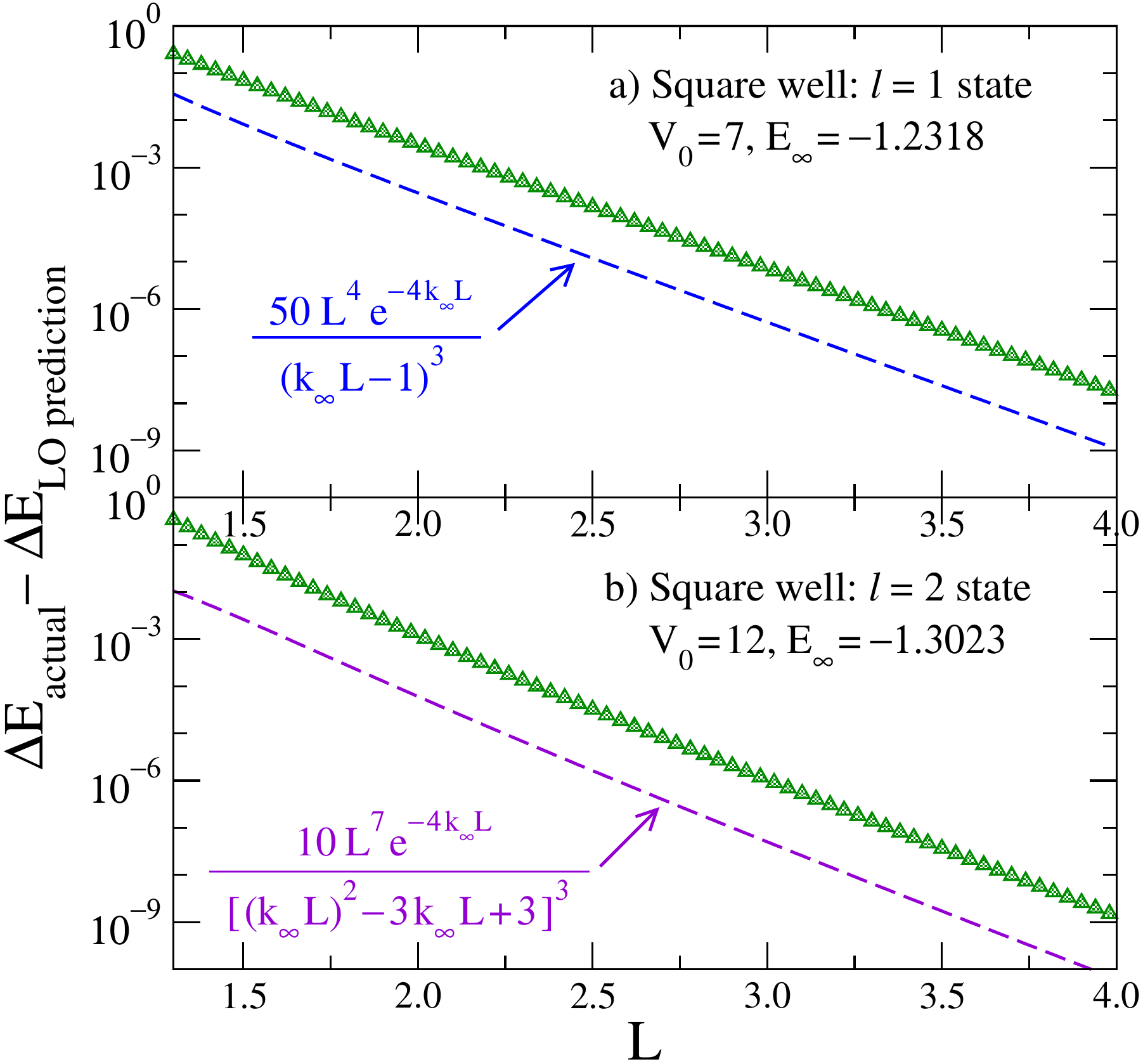}
\caption{(color online) Error plots of the energy correction at each
  $L$ for a) $l=1$ and b) $l=2$ square-well states predicted at leading order by
  Eqs.~\eqref{eq:DeltaE_l1} and \eqref{eq:DeltaE_l2} compared to the exact energy.
  Lines proportional to the expected L-NLO residual errors   
  are plotted for comparison. }
\label{fig:sq_well_error_log_plots_P_and_D}
\end{figure}

In complete analogy to the case of $s$ waves (e.g., using
\eqref{eq:basiceq} and \eqref{eq:purepole} for general $l$), the
correction $\Delta E$ of the energy at leading order is
\beq
[\Delta E]_{\rm LO} = - \kinf \left(\gamma_\infty^{(l)}\right)^2 
{h_l^{(1)}(ik_L L)\over h_l^{(1)}(-ik_L L)} \ .
  \label{eq:ELO_general_l}
\eeq
We note that 
\beq
{h_l^{(1)}(ix)\over h_l^{(1)}(-ix)} \approx - e^{-2x}
\eeq
for $x\gg 1$. In particular, for $l=1$
\beq
[\Delta E]_{\rm LO} =\kinf \left(\gamma_\infty^{(1)}\right)^2
{\kinf L+1\over \kinf L-1} \, e^{-2\kinf L}  \;,
\label{eq:DeltaE_l1}
\eeq
and for $l=2$
\beq
[\Delta E]_{\rm LO} =\kinf \left(\gamma_\infty^{(2)}\right)^2
{(\kinf L)^2 +3 \kinf L +3\over (\kinf L)^2-3\kinf L +3}\, e^{-2\kinf L} \;.
  \label{eq:DeltaE_l2}
\eeq
These correction formulas are tested in Fig.~\ref{fig:sq_well_error_log_plots_P_and_D}.
For coupled channels, the leading energy correction will be the sum
of the LO corrections for the individual angular momenta.
We note that lattices with periodic bc lead to energy shifts that
depend on the angular momentum~\cite{Konig:2011nz}.  
In contrast, the basis truncations we consider in this work are
variational and thus always yield a positive energy correction.

\begin{figure}[thb!]
\centering
\includegraphics[width=0.95\columnwidth]{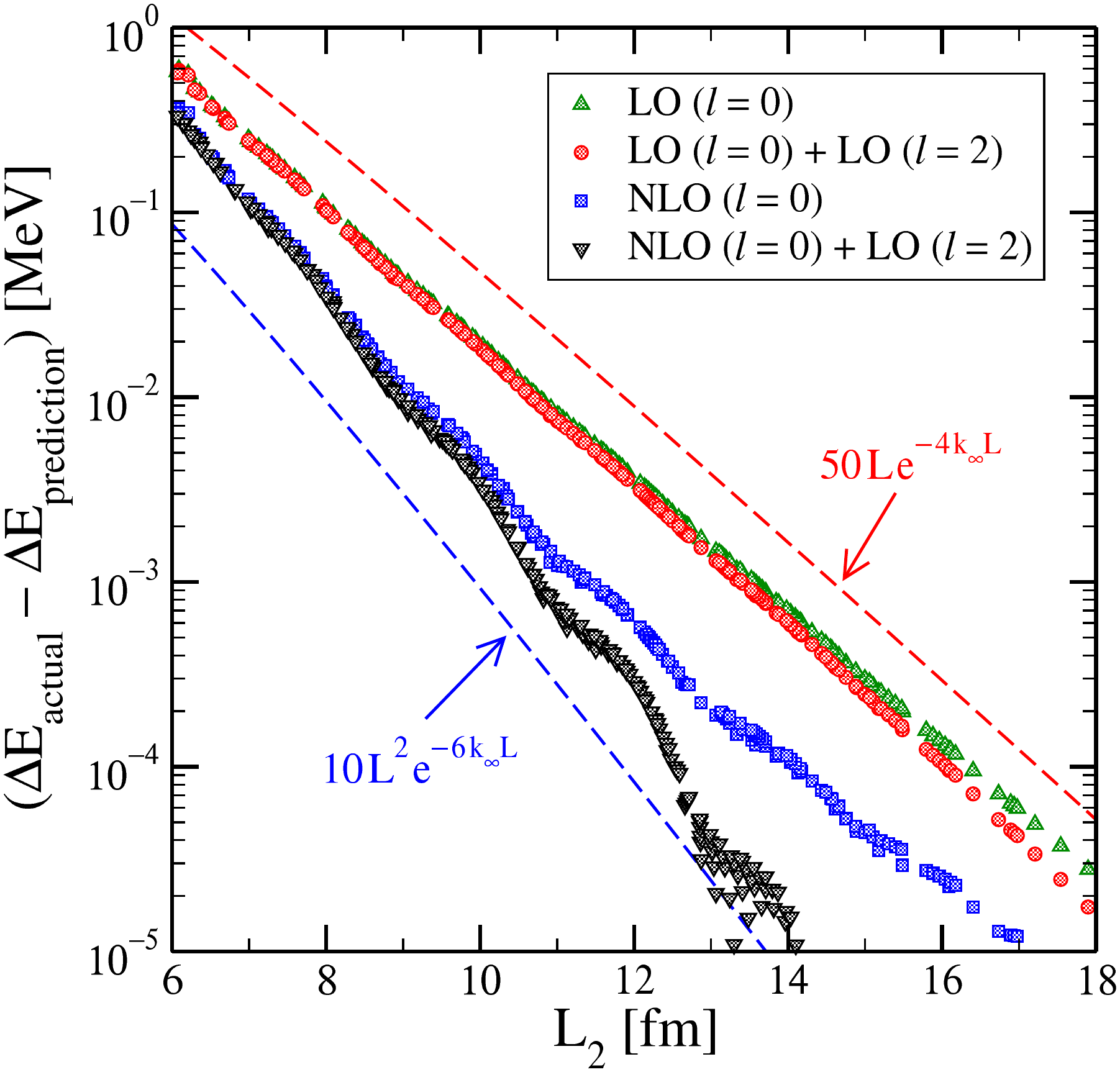}
\caption{(color online) Residual error for the deuteron energy due to
  HO basis truncation as a function of $L = L_2$ (with $\hw > 49$~MeV
  to eliminate UV contamination) after subtracting $l=0$ energy
  corrections at different orders from
  Eq.~\eqref{eq:complete_E_correction_NLO} and the $l=2$ correction
  from Eq.~\eqref{eq:DeltaE_l2}.  For the parameter $w_2$ in
  Eq.~\eqref{eq:complete_E_correction_NLO} we use the value reported
  in \cite{Phillips:1999hh}.  }
\label{fig:deuteron_error_plot_l2}
\end{figure}

We return to the deuteron and take 
$|\gamma_\infty^{(2)}/\gamma_\infty^{(0)}| \approx 0.0226/0.8843$ from 
Ref.~\cite{Machleidt:2011zz}. Then
\bea
\lefteqn{[\Delta E]_{\rm LO} =\kinf \left(\gamma_\infty^{(0)}\right)^2 
e^{-2\kinf L}}
  \nonumber \\
  & \qquad \null \times
\left[ 1 + 
\left|\frac{\gamma_\infty^{(2)}}{\gamma_\infty^{(0)}}\right|^2
\frac{(\kinf L)^2 +3 \kinf L +3}{(\kinf L)^2-3\kinf L +3}\right]
\;.
\eea
This formula is tested in Fig.~\ref{fig:deuteron_error_plot_l2} with
the same deuteron calculations as in
Fig.~\ref{fig:deuteron_error_log_plots}. We note that the deviation
after subtraction of the NLO ($l=0$) result does not exhibit the
$\exp(-6k_\infty L)$ falloff but is rather consistent with an
$\exp(-4\kinf L)$ falloff at large $L$.  We attribute this to the
missing LO $d$-state correction. Due to the small value of the
$d$-to-$s$ state ratio, the $d$-wave correction is small, but it makes
a perceptible shift of the $s$-wave LO result.  When added to the NLO
$l=0$ correction, the large $L_2$ behavior of the error is brought
somewhat closer in line with the predicted dependence of $L^2
e^{-6\kinf L}$. We note, however, that the NLO correction is not complete
due to the missing $l=2$ correction.


\section{Alternative methods}\label{sec:derivative_method}

In this Section we briefly consider two alternative approaches to the
expansion for $\Delta E_L$.  The linear energy method~\cite{Djajaputra:2000aa} was used in
Refs.~\cite{Furnstahl:2012qg,More:2013rma} to derive the form
of the expansion and the leading term.  A modified correction to LO
for shallow bound states was also suggested in
Ref.~\cite{More:2013rma}, but we have found that it is not part of a
consistent expansion; we correct it here.  The other method constructs
the differential variation of the energy with $L$, which can be
integrated to reproduce our present expansion.

\subsection{Linear energy method}

The linear energy method is based on the observation that the regular
radial solution $u_E(r)$ for energy $E$ has a smooth expansion about
$E = \Einf$ at fixed $r$, so that the wave function for $E=E_L$, which
is denoted $u_L(r)$, can be approximated by
\beq
  u_L(r)\approx u_\infty (r) + \Delta E_L
  \left.\frac{du_E(r)}{dE}\right|_{E_{\infty}}
  + \mathcal{O}(\Delta E_L^2) \;,
  \label{eq:linear_energy_approx}
\eeq
for $r \leq L$.  By evaluating at $r=L$ with the boundary condition
$u_L(L)=0$, $\Delta E_L$ is estimated as
\beq
  \Delta E_L \approx -u_\infty(L) \left(\left.\frac{d u_E(L)}
  {dE}\right|_{\Einf}\right)^{-1}
  \;.
  \label{eq:Delta_EL_linear}
\eeq
The leading approximation to $du_E(L)/dE|_{\Einf}$ then leads to the
LO result $\Delta E_L \propto e^{-2\kinf L}$ with the coefficient
correctly identified in Ref.~\cite{More:2013rma}.

The modified energy correction proposed in Ref.~\cite{More:2013rma},
\beq
 (\Delta E_L)_{\rm mod} = \kinf \ANC ^2
   \frac{e^{-2 \kinf L}}{(1-\ANC^2 L e^{-2 \kinf L})}
   \;,
  \label{eq:IR_mod_scaling}
\eeq
contains all orders in the expansion factor $e^{-2\kinf L}$. However, if
expanded in a power series it does not reproduce correctly the
$L$-dependent $e^{-4\kinf L}$ term in
Eq.~\eqref{eq:complete_E_correction_NLO}.  In light of the consistent
expansion presented in this paper, it is clear that a term of
$\mathcal{O}(L\ANC^2e^{-4\kinf L})$ also arises from the
$\mathcal{O}(\Delta E_L^2)$ term in
Eq.~\eqref{eq:linear_energy_approx}. When this contribution is taken
into account the result from $(\Delta E_L)_{\rm mod}$ matches that
from Eqs.~\eqref{eq:complete_E_correction_NLO} up to \LNLO.

\subsection{Differential method}

Because we seek the change in energy with respect to a cutoff, it is
natural to formulate the problem in the spirit of renormalization
group methods by seeking a flow equation for the bound-state energy as
a function of $L$.  Such an approach is already documented in the
literature, for example in Refs.~\cite{Arteca1984} and
\cite{Fernandez1981}, and it provides us with an alternative method that
does not directly reference the S-matrix.  The basic equation is
\beq
  \frac{\partial E_L}{\partial L} = -\frac12 \frac{|u'_L(L)|^2}{\int_0^L |u_L(r)|^2\, dr}
  \;.
  \label{eq:dEdL}
\eeq
Here the prime denotes a derivative with respect to $r$.  Given an
expression for the right-hand side in terms of observables ($\kinf$,
$\ANC$, and so on) and $L$, we can simply integrate to find the energy
correction for a bc at $L$
\beq
  \Delta E_L \equiv E_L - \Einf = \int_{\Einf}^{E_L}\! dE\,
     = \int_\infty^L\! \frac{\partial E_L}{\partial L} dL
     \;.
\eeq
To derive Eq.~\eqref{eq:dEdL}, we start with
\beq
  \frac{\partial}{\partial L}\left [
   \int_0^L u_L(r) H u_L(r)\, dr = E_L \int_0^L\! dr\, u_L(r)^2
  \right]
  \;,
\eeq
which yields (after some cancellations)
\beq
    \frac12 \left.\left( \frac{\partial u_L(r)}{\partial r} 
         \frac{\partial u_L(r)}{\partial L}
         \right)\right|_0^L
   =  
  \frac{\partial E_L}{\partial L} \int_0^L\! dr\, u_L(r)^2
         \;.
         \label{eq:DeltaEL}
\eeq
The left-hand side is a surface term from partially integrating the
kinetic energy in $H$.  The lower limit vanishes because $u_L(0) = 0$
for any $L$.  Finally, we replace the partial derivative with respect
to $L$ at the upper limit using
\beq
   \frac{\partial u_L(L)}{\partial L} = - \frac{\partial u_L(L)}{\partial r}
   \;,
\eeq
which follows from expanding $u_{L'}(L') = 0$ about $u_{L}(L) = 0$ for $L' = L + \Delta L$.

To apply Eq.~\eqref{eq:dEdL}, we start with $u_L(r)$ in the asymptotic
region, as given by Eq.~\eqref{eq:uLasymp2}.  The normalization
constant $\gamma_L$ is chosen so that the integral of $u_L(r)^2$ from
0 to $L$ is unity; it becomes the ANC $\ANC$ as
$L\rightarrow\infty$.  Thus
\beq
   u_L'(L) = -2 \gamma_L k_L e^{-k_L L}
   \;.
   \label{eq:uLprime}
\eeq
Now we need to expand $k_L$ and $\gamma_L$ about $\kinf$ and $\ANC$,
respectively.  The leading term is trivial: $k_L \rightarrow \kinf$
and $\gamma_L \rightarrow \ANC$, so the only $L$ dependence in
$u_L'(L)^2$ is in $e^{-2 \kinf L}$ and the integration in
\eqref{eq:dEdL} is immediate:
\begin{align}
\Delta E_L
     &= \int_\infty^L\! \frac{\partial E_L}{\partial L} dL
     =  -2 \ANC^2 \kinf^2 \int_\infty^L\! e^{-2\kinf L} \, dL
     \nonumber \\
     &=  \kinf \ANC^2 e^{-2 \kinf L} +  \mathcal{O}(e^{-4 \kinf L})
     \;.
\end{align}
This is the same LO result for $\Delta E_L$ found by other methods.
It is straightforward to extend this construction to $l>0$, reproducing
Eq.~\eqref{eq:ELO_general_l}.

To go to NLO we need an expression for $\gamma_L$.  In the zero-range
(zr) limit, $\gamma_L$ is given completely in terms of $k_L$ using the
normalization condition (because the asymptotic form in
Eq.~\eqref{eq:uLasymp2} holds over the entire range of the integral)
\begin{align}
  \gamma_L^2 &= \left[\int_0^L\! dr\, (e^{-k_L r} - e^{-2k_L L}e^{k_L r})^2 \right]^{-1}
  \nonumber \\ &=
  2 k_L(1 + 4 k_L L e^{-2 k_L L}) + \mathcal{O}(e^{-4 k_L L}) \;.
\end{align}
We expand $k_L$ everywhere in Eq.~\eqref{eq:dEdL} using
Eq.~\eqref{eq:uLprime} and our LO result
\beq
   k_L = \kinf (1 - 2 e^{-2 \kinf L})
   \;.
\eeq
Here, we neglected terms that are $\mathcal{O}(e^{-6 \kinf L})$ or smaller.
We need to expand $e^{-2 k_L L}$ in $u_L'(L)$ to get
\beq
   e^{-2 k_L L} = e^{-2 \kinf L} (1 + 4 \kinf L e^{-2 \kinf L}) 
    + \mathcal{O}(e^{-6 \kinf L}) \;.
\eeq
(Elsewhere it suffices to replace $e^{-2 k_L L}$ by $e^{-2 \kinf L}$ to NLO.)
So we find that
\begin{align}
  \frac{\partial E_L}{\partial L} &= -\frac12 (4 \gamma_L^2 k_L^2 e^{-2 k_L L})
   \nonumber \\
     &\approx -2  [2\kinf (1 - 2 e^{-2 \kinf L})(1 + 4 \kinf L e^{-2 \kinf L})]
     \nonumber \\
     & \  \null \times [\kinf^2 (1 - 4 e^{-2\kinf L})][e^{-2 \kinf L} (1 + 4 \kinf L e^{-2 \kinf L})]
    \nonumber \\
    &\approx
    -4 \kinf^3 e^{-2 \kinf L} -  8 \kinf^3 (4\kinf L - 3)e^{-4 \kinf L} 
      \nonumber \\ & \qquad \null + \mathcal{O}(e^{-6 \kinf L})
    \;,
\end{align}
and then finally
\begin{align}
[\Delta E_L]_{\rm zr, NLO}
     &= \int_\infty^L\! \frac{\partial E_L}{\partial L} dL
     \nonumber \\
     &=
     2 \kinf^2 e^{-2 \kinf L} +  4 \kinf^2 (2\kinf L - 1)e^{-4 \kinf L} 
     \nonumber \\
     & \qquad \null + \mathcal{O}(e^{-6 \kinf L})
     \;,
\end{align}
in agreement with Eq.~\eqref{eq:complete_E_correction_NLO} 
with $\ANC^2 = 2\kinf$ and $w_2 = 0$.
We can take this procedure to higher order by
using a more general expansion for $k_L$.

To extend the differential method
to higher order for nonzero range, we must parametrize
$\gamma_L$ to account for the part of the integration within the
range of the potential; e.g., in terms of the effective range.  However,
we have not found a clear advantage in doing this
compared to the straightforward S-matrix method.

\section{Radii} \label{sec:radii}

In this Section, we compute corrections to the radius for $l=0$ 
to $\mathcal{O}(e^{-2\kinf L})$. 
The corresponding formula was given in
Ref.~\cite{Furnstahl:2012qg} without a derivation. 
We define 
\beq
  \rsqav_L = \rsqinf + \Delta\rsqav_L
  \;,
\eeq
where
\beq
  \Delta\rsqav_L = 
  \frac{\int_0^L |u_L(r)|^2\, r^2\,dr}{\int_0^L |u_L(r)|^2\, dr}
   - \frac{\int_0^\infty |u_\infty(r)|^2\, r^2\,dr}{\int_0^\infty |u_\infty(r)|^2\, dr}
   \;.
   \label{eq:Delta-rsq}
\eeq   
Though the squared radius is a long-ranged
operator, its matrix elements will still be modified at short distances
by renormalizations or  similarity transformations of the Hamiltonian, 
see, e.g., Ref.~\cite{Stetcu:2004wh}.  Thus we cannot expect an extrapolation 
law for the radius that depends entirely on observables.
Instead, we seek a formula that identifies the $L$ dependence but leaves
parameters to be fit.

The strategy is to isolate the polynomial $L$ dependence by splitting the
necessary integrals into an interior part and an exterior part:
\beq
  \int_0^L\! r^n |u_L(r)|^2\, dr =
  \int_0^R\! r^n |u_L(r)|^2\, dr + \int_R^L\! r^n |u_L(r)|^2\, dr
  \;,
  \label{eq:rnintegral}
\eeq
where $R$ is sufficiently large so that the asymptotic form of $u_L(r)$ from
Eq.~\eqref{eq:uLasymp2} can be used in the second integral.
Our expression for $\Delta\rsqav_L$ is independent of the normalization
of $u_L(r)$, so we are free to choose it so that the large $r$ form is
exactly given by Eq.~\eqref{eq:uLasymp2}.

The first integral will depend on the details of the interior wave function
and therefore on the potential, but 
the linear energy method shows us that to $\mathcal{O}(e^{-2\kinf L})$
the $L$ dependence is isolated.  In particular, the dependence on $L$ of $u_L(r)$ in
Eq.~\eqref{eq:linear_energy_approx} is confined to 
$\Delta E_L = \kinf\ANC^2 e^{-2\kinf L}$ because $du_E(r)/dE|_{\Einf}$ for
$r < R$ is independent of $L$ with our choice of normalization.  Thus the
integral over $r$ cannot introduce polynomial $L$ dependence and we can
conclude that
\beq
  \int_0^R\! r^n |u_L(r)|^2\, dr = \mathcal{O}(L^0) e^{-2\kinf L}
     + \mathcal{O}(e^{-4\kinf L}) \;. 
\eeq
The $\mathcal{O}(L^0)$ coefficient will depend on the potential, so we
will treat it as a parameter to be fit.

The second integral can be directly evaluated to $\mathcal{O}(e^{-2\kinf L})$
using Eq.~\eqref{eq:uLasymp2} and $[k_L]_{LO} = \kinf - \ANC^2 e^{-2\kinf L}$
to expand $|u_L(r)|^2$.  For $n=0$ we find
\begin{align}
  \int_R^L\! &|u_L(r)|^2\, dr =
  \frac{1}{2\kinf}e^{-2\kinf R}
   \nonumber \\
  & \null + \Bigl[
   \frac{\ANC^2}{\kinf}
      \Bigl(R + \frac{1}{2\kinf}\Bigr)e^{-2\kinf R}
     +2R - 2L
  \Bigr]e^{-2\kinf L}
     \nonumber \\
  & \null + \mathcal{O}(e^{-4\kinf L})
  \;,
  \label{eq:asympnorm}
\end{align}
and for $n=2$ we find
\begin{align}
  \int_R^L\! &r^2 |u_L(r)|^2\, dr =
  \frac{1}{2\kinf^3}\Bigl[ \frac12 + \kinf R + (\kinf R)^2 \Bigr] e^{-2\kinf R}
   \nonumber \\
  & \null + 
  \Bigl[
   \frac{\ANC^2}{\kinf^4}  
       \Bigl(
           \frac34 + \frac32\kinf R + \frac32(\kinf R)^2 + (\kinf R)^3  
       \Bigr)e^{-2\kinf R}
   \nonumber \\
   & \qquad \null 
   + \frac{1}{\kinf^3}  
       \Bigl(
         \frac23 (\kinf R)^3 - \kinf L - \frac23 (\kinf L)^3
       \Bigr)
  \Bigr]e^{-2\kinf L}
     \nonumber \\
  & \null + \mathcal{O}(e^{-4\kinf L})
  \;.
    \label{eq:asymprsq}
\end{align}
Note that it is necessary to keep the expansion of $|u_L(r)|^2$ up to $e^{-4\kinf L}$ 
until after doing the integrals because terms proportional to  
$e^{-4\kinf L} e^{2\kinf r}$ will be leading order.

When we use \eqref{eq:asympnorm} and \eqref{eq:asymprsq} and our previous
result for the interior integrals in Eq.~\eqref{eq:Delta-rsq}, expanding
consistently to $\mathcal{O}(e^{-2\kinf L})$, we will mix $R$-dependent
terms with the $L$ dependence.  However, we can immediately conclude that
the general form to this order is (with $\beta \equiv 2\kinf L$)
\bea
 \rsqav_L \approx {\rsqinf}[ 1 - (c_0 \beta^3 +c_1 \beta + c_2) e^{-\beta}]
   \;.
\label{rad}
\eea
Here, $\rsqinf$, $c_0$, $c_1$, and $c_2$ are fit parameters while $\kinf$
should be determined from fitting the energy. 
This form has been verified explicitly for finite-range model potentials (e.g., square well
and delta shell).
The approximation~(\ref{rad}) should be valid
in the asymptotic regime $\beta\gg 1$. In practice, for a robust
extrapolation one needs
$\beta$ large enough so that the $\beta^3$ correction 
dominates the subleading terms.

\begin{figure}[tbh!]
\centering
\includegraphics[width=0.95\columnwidth]{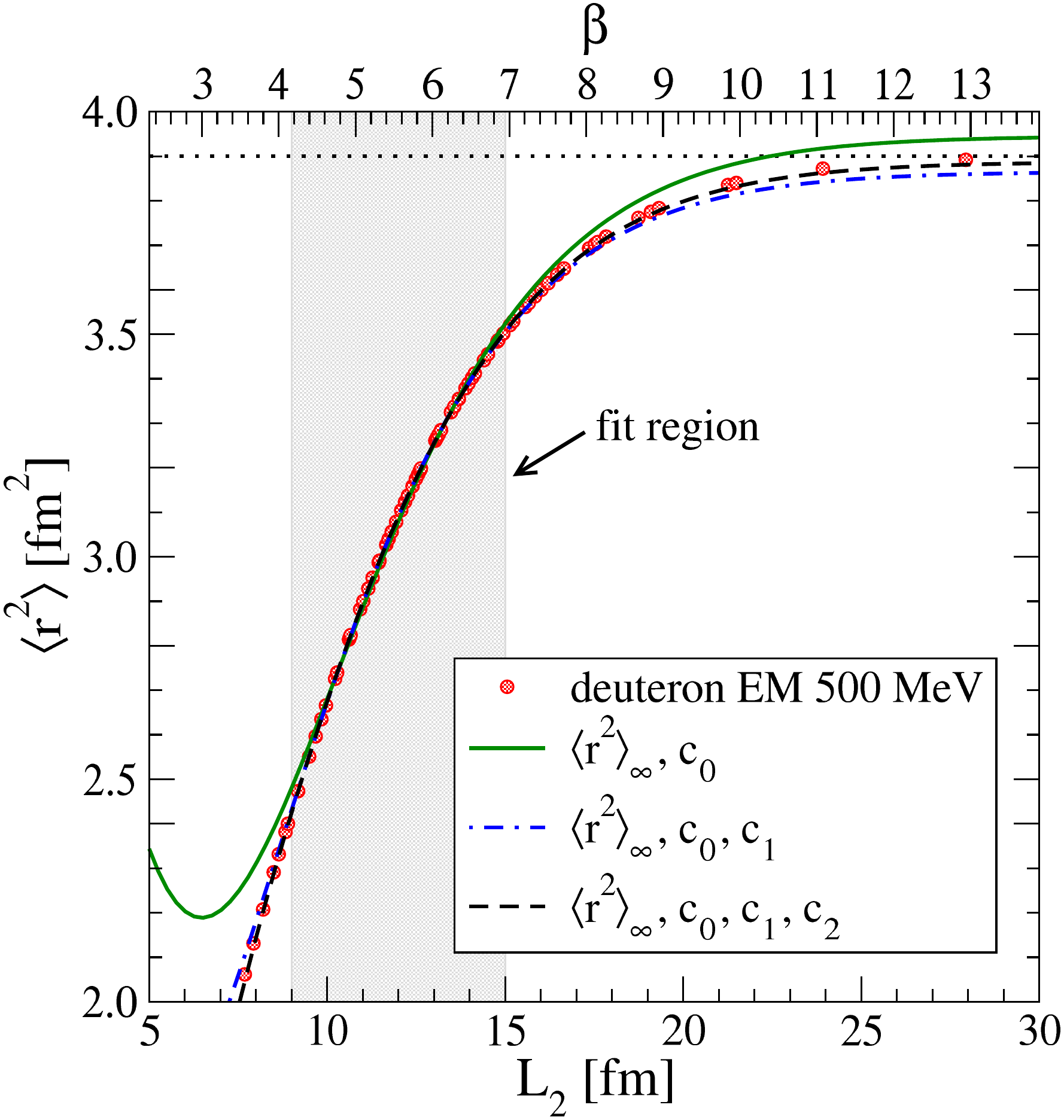}
\caption{(color online) Deuteron radius squared versus $L_2$ 
  for the chiral N$^3$LO (500\,MeV) potential
  of Ref.~\cite{Entem:2003ft}.  To eliminate the UV contamination we
  only plot results for $\hw > 49$~MeV.  The solid, dot-dashed, and dashed
  lines are results from fitting Eq.~\eqref{rad} in the shaded
  region to find
  $\rsqinf$ and one, two, or three of the $c_i$ constants, respectively.  
  The horizontal dotted line is the deuteron radius squared.}
\label{fig:deuteron_radius_fit}
\end{figure}

If we take the zero-range limit $R\to 0$ of the potential, we 
arrive at the simple expression
\beq
 \frac{\Delta\rsqav_L}{\rsqinf} 
   \approx
    -\left(\frac{(2\kinf L)^3}{3} -4\right) e^{-2\kinf L}
   \;.
   \label{eq:rsqzerorange}
\eeq
Note that in this limit the correction becomes independent of the potential.
Equation~\eqref{eq:rsqzerorange} suggests that for a short-range potential,
the $c_1$ and $c_2$ terms will give comparable contributions for moderate
$\beta$, and therefore will be difficult to determine reliably.

Sample fits of Eq.~\eqref{rad} for the deuteron are shown in 
Fig.~\ref{fig:deuteron_radius_fit}. Results are given for fitting
one, two, and all three $c_i$ constants to radii calculated
with the same truncated oscillator basis parameters used for 
Fig.~\ref{fig:deuteron_curves}.  The fit region is for $L_2$
between 9 and 15\,fm, where the calculations only show a small
amount of curvature.  All points are equally weighted.
The extrapolated radius squared $\rsqinf$ for 
the three cases are 3.95\,fm$^2$, 3.87\,fm$^2$, and 3.89\,fm$^2$,
compared to the exact result of 3.90\,fm$^2$.  
If the fit region is instead taken between 11 and 17\,fm, the
corresponding results are 3.91\,fm$^2$, 3.89\,fm$^2$, and 
3.90\,fm$^2$.  
For all of these fits, the value of $c_0$ is
fairly stable, ranging from 0.27 to 0.33 (note that
$c_0 = 1/3$ in the zero-range limit).  In contrast, $c_1$
and $c_2$ are not well determined (even the sign of $c_1$
varies).  This is consistent with
fits using the square-well potential, where analytic expressions
for the $c_i$s can be found.  We find that $\rsqinf$
and $c_0$ are well determined by fits in analogous
regions but that $c_1$ and
$c_2$ are not.
If we push the analysis by taking the fit region between 7 and 13\,fm, 
the $\rsqinf$ prediction using only $c_0$ breaks down, giving
4.21\,fm$^2$.  However, the fit with all three $c_i$s is still
reasonable, giving 3.86\,fm$^2$. 
Further studies are needed to test how these trends might carry
over to $A>2$ nuclei.

The derivation given here can be directly extended to $l>0$ using the general
expression for the asymptotic wave function in Eq.~\eqref{eq:uLasympl}.
However, this wave function has additional $L$ dependence so the corresponding
result to Eq.~\eqref{rad} will have more complicated $\beta$ dependence
unless additional simplifications are made.  The extension to other single-particle
coordinate-space operators is also direct, by replacing $r^2$ with the
appropriate expression.


\section{Summary and open questions}\label{sec:summary}

In this paper we derived and tested a consistent and systematic
expansion for the $s$-wave binding momentum and energy of a two-body
system with a Dirichlet boundary condition,
Eqs.~\eqref{eq:complete_k_correction_NLO} and
\eqref{eq:complete_E_correction_NLO}.  As shown in
Ref.~\cite{More:2013rma} for $l=0$ bound states, such a boundary
condition arises as an effective infrared cutoff when using a
truncated harmonic oscillator basis.  Here we extended to $l>0$ the
derivation from \cite{More:2013rma} that associates the oscillator
basis parameters to the appropriate hard-wall radius $L$. The same
formula for $L$ derived previously for $l=0$ (called $L_2$) is found
to still hold for general $l$ if expressed in terms of the oscillator
quantum number $N = 2n + l$.  We subsequently obtained the energy
correction for $l>0$ at LO.

Our expansion is based on the analytic structure of the two-body
S-matrix in the complex momentum plane.  The asymptotic wave functions
for a boundary condition at $r=L$ are analytic continuations of the
scattering solutions to (purely) imaginary momentum.  If continued to
$k = i\kinf$, the free-space $L=\infty$ binding momentum, one reaches
a pole of the S-matrix with residue determined by the asymptotic
normalization $\ANC$.  If there are no long-range interactions that
generate intermediate singularities, as is the case for the deuteron
where the one-pion exchange threshold is further away, this entire
continuation is determined by measurable quantities (the on-shell
S-matrix).  The binding momentum $k_L$ for the boundary condition at
$L$ is intermediate between zero and $\kinf$ and therefore it is
determined by observables.

The expansion for $k_L$ and subsequently $E_L$ is naturally formulated
using an effective range expansion of $k\cot\delta_l(k)$ about the
pole at $i\kinf$.  The expansion is in powers of $e^{-2\kinf L}$ (LO
goes like $e^{-2\kinf L}$, NLO like $e^{-4\kinf L}$, and so on), with
prefactors that depend on $L$, $\kinf$, $\ANC$, and higher-order
effective range parameters.  The leading term and the $L$-dependent
NLO term are determined by the pole alone, while other NLO and
higher-order terms require a valid parameterization of the S-matrix
away from the pole.  (For a zero-range interaction, the expansion
depends on $\kinf$ only.)  This organization was tested for
model potentials (not all shown here)  
and a realistic deuteron calculation (the latter
within a harmonic oscillator basis).  The use of semi-log error plots
to compare to analytic results for the square well and to
numerical results for the deuteron demonstrates the validity of the expansion
over a wide range of $L$.  We found the use of error plots to be a
much more robust test than simply graphing the approach to $\Einf$.

The use of a Dirichlet boundary condition is only an approximation to
the actual asymptotic behavior of the harmonic oscillator basis in
coordinate representation.  
However, as illustrated in Fig.~\ref{fig:HO_q_and_rspace_sqwell},
the difference in behavior is a high-momentum effect.
Our error plots for the deuteron, which represent an
indirect comparison because the energies were found from
oscillator-truncated diagonalizations, suggest the former corrections
remain subleading to the NLO corrections 
for $E_L$ over a wide range in $L$.  We note that LO
corrections due to the $d$-wave component of the deuteron are small
but push the error plots to good agreement with the expected LO
error proportional to $L e^{-4\kinf L}$ at the largest $L$ values.

Two alternative derivations of the expansions for $k_L$ and $E_L$ were
also presented, based on the linear energy method and a differential
method, respectively.  For the former, we corrected the modified
version of the expansion for $E_L$ proposed in \cite{More:2013rma},
which was not consistent at $\mathcal{O}(e^{-4\kinf L})$.  While these
alternatives provide different perspectives on the energy corrections,
we did not find any new capabilities thus far.  However, they may be
more useful in calculations of other quantities (which can be scheme
dependent); this is being explored.

The formulation in terms of S-matrix analytic structure is closely
related to methods used to analyze break-up reactions, which provides a
link to $A>2$ extrapolations.  Indeed, in Ref.~\cite{Furnstahl:2012qg}
the basic form of the LO extrapolation proportional to $e^{-2\kinf L}$
was based on interpreting $\kinf$ in terms of the one-particle
separation energy.  More generally, the asymptotic many-body wave
function is dominated by configurations corresponding to the break-up
channels with the lowest separation energies and it is
their modification by the hard wall at $L$ that will be associated
with the energy shift $\Delta E_L$.  This is in turn dominantly
described by the S-matrix near poles at the corresponding separation
binding momenta.  Future work will seek to clarify the precise nature
of the more general expansion (including the effects of the Coulomb
interaction) and whether it will be possible to quantitatively extract
asymptotic normalization constants.

It might be challenging to derive NLO corrections to the binding
energies for nuclei with $A> 2$, particular for nuclei with nonzero
ground-state spin. Here, many different orbital anglar momenta can
contribute to the ground-state wave function, and one would presumably
need to know the admixture of the different channels quite accurately.
Our results show that nonzero orbital angular momenta yield
corrections in inverse powers of $\kinf L$ to the LO energy
extrapolation. On the other hand, the leading contributions to
bound-state energies in finite model spaces fall off as $\exp{(-2\kinf
  L)}$ for all orbital angular momenta. This makes extrapolations
feasible in practice.

Corrections due to
the UV cutoff induced by a finite oscillator space were not considered
in the present work, because the effective oscillator momentum
cutoffs used (e.g., for the deuteron in Figs.~\ref{fig:deuteron_curves},
\ref{fig:deuteron_error_log_plots}, and \ref{fig:deuteron_error_plot_l2})
were well above the intrinsic cutoff of the input potential.
We have demonstrated that in this case the energy will be converged
in the UV.  However, rendering the UV correction negligible may not
always be practical for larger systems with some
methods~\cite{Roth:2013fqa}.  This motivates, together with the success
of phenomenological extrapolation schemes, a search for theoretically
founded schemes that combine IR and UV expansions.
Our systematic IR expansion relies on the IR cutoff being in the
asymptotic region in coordinate space, beyond the range of the
potential. The UV cutoff is at high momentum, however, where the potential is
directly modified.  While the duality of the oscillator Hamiltonian
implies that the UV cutoff will be well approximated by a hard cutoff
at a momentum given by the analogous expression to Eq.~\eqref{eq:L2_def},
the energy corrections will be dependent on the potential (i.e., not dependent
only on observables).  These issues will be explored in a forthcoming
publication.


\medskip

\begin{acknowledgments}
  We thank B.~Dainton, H.~Hergert, S.~Koenig, and R.~Perry  for useful
  discussions.  This work was supported in part by the National
  Science Foundation under Grant No.~PHY--1002478 and the Department
  of Energy under Grant Nos.~DE-FG02-96ER40963 (University of
  Tennessee), DE-AC05-00OR22725 (Oak Ridge National Laboratory), and
  DE-SC0008499/DE-SC0008533 (SciDAC-3 NUCLEI project).
\end{acknowledgments}

\bibliography{srg_refs}

\begin{thebibliography}{35}%
\makeatletter
\providecommand \@ifxundefined [1]{%
 \@ifx{#1\undefined}
}%
\providecommand \@ifnum [1]{%
 \ifnum #1\expandafter \@firstoftwo
 \else \expandafter \@secondoftwo
 \fi
}%
\providecommand \@ifx [1]{%
 \ifx #1\expandafter \@firstoftwo
 \else \expandafter \@secondoftwo
 \fi
}%
\providecommand \natexlab [1]{#1}%
\providecommand \enquote  [1]{``#1''}%
\providecommand \bibnamefont  [1]{#1}%
\providecommand \bibfnamefont [1]{#1}%
\providecommand \citenamefont [1]{#1}%
\providecommand \href@noop [0]{\@secondoftwo}%
\providecommand \href [0]{\begingroup \@sanitize@url \@href}%
\providecommand \@href[1]{\@@startlink{#1}\@@href}%
\providecommand \@@href[1]{\endgroup#1\@@endlink}%
\providecommand \@sanitize@url [0]{\catcode `\\12\catcode `\$12\catcode
  `\&12\catcode `\#12\catcode `\^12\catcode `\_12\catcode `\%12\relax}%
\providecommand \@@startlink[1]{}%
\providecommand \@@endlink[0]{}%
\providecommand \url  [0]{\begingroup\@sanitize@url \@url }%
\providecommand \@url [1]{\endgroup\@href {#1}{\urlprefix }}%
\providecommand \urlprefix  [0]{URL }%
\providecommand \Eprint [0]{\href }%
\providecommand \doibase [0]{http://dx.doi.org/}%
\providecommand \selectlanguage [0]{\@gobble}%
\providecommand \bibinfo  [0]{\@secondoftwo}%
\providecommand \bibfield  [0]{\@secondoftwo}%
\providecommand \translation [1]{[#1]}%
\providecommand \BibitemOpen [0]{}%
\providecommand \bibitemStop [0]{}%
\providecommand \bibitemNoStop [0]{.\EOS\space}%
\providecommand \EOS [0]{\spacefactor3000\relax}%
\providecommand \BibitemShut  [1]{\csname bibitem#1\endcsname}%
\let\auto@bib@innerbib\@empty
\bibitem [{\citenamefont {Stetcu}\ \emph {et~al.}(2007)\citenamefont {Stetcu},
  \citenamefont {Barrett},\ and\ \citenamefont {van Kolck}}]{Stetcu:2006ey}%
  \BibitemOpen
  \bibfield  {author} {\bibinfo {author} {\bibfnamefont {I.}~\bibnamefont
  {Stetcu}}, \bibinfo {author} {\bibfnamefont {B.~R.}\ \bibnamefont {Barrett}},
  \ and\ \bibinfo {author} {\bibfnamefont {U.}~\bibnamefont {van Kolck}},\
  }\href@noop {} {\bibfield  {journal} {\bibinfo  {journal} {Phys. Lett. B}\
  }\textbf {\bibinfo {volume} {653}},\ \bibinfo {pages} {358} (\bibinfo {year}
  {2007})}\BibitemShut {NoStop}%
\bibitem [{\citenamefont {Hagen}\ \emph {et~al.}(2010)\citenamefont {Hagen},
  \citenamefont {Papenbrock}, \citenamefont {Dean},\ and\ \citenamefont
  {Hjorth-Jensen}}]{hagen:2010gd}%
  \BibitemOpen
  \bibfield  {author} {\bibinfo {author} {\bibfnamefont {G.}~\bibnamefont
  {Hagen}}, \bibinfo {author} {\bibfnamefont {T.}~\bibnamefont {Papenbrock}},
  \bibinfo {author} {\bibfnamefont {D.}~\bibnamefont {Dean}}, \ and\ \bibinfo
  {author} {\bibfnamefont {M.}~\bibnamefont {Hjorth-Jensen}},\ }\href {\doibase
  10.1103/PhysRevC.82.034330} {\bibfield  {journal} {\bibinfo  {journal} {Phys.
  Rev. C}\ }\textbf {\bibinfo {volume} {82}},\ \bibinfo {pages} {034330}
  (\bibinfo {year} {2010})}\BibitemShut {NoStop}%
\bibitem [{\citenamefont {Jurgenson}\ \emph {et~al.}(2011)\citenamefont
  {Jurgenson}, \citenamefont {Navratil},\ and\ \citenamefont
  {Furnstahl}}]{jurgenson:2010wy}%
  \BibitemOpen
  \bibfield  {author} {\bibinfo {author} {\bibfnamefont {E.~D.}\ \bibnamefont
  {Jurgenson}}, \bibinfo {author} {\bibfnamefont {P.}~\bibnamefont {Navratil}},
  \ and\ \bibinfo {author} {\bibfnamefont {R.~J.}\ \bibnamefont {Furnstahl}},\
  }\href {\doibase 10.1103/PhysRevC.83.034301} {\bibfield  {journal} {\bibinfo
  {journal} {Phys. Rev. C}\ }\textbf {\bibinfo {volume} {83}},\ \bibinfo
  {pages} {034301} (\bibinfo {year} {2011})}\BibitemShut {NoStop}%
\bibitem [{\citenamefont {Coon}\ \emph {et~al.}(2012)\citenamefont {Coon},
  \citenamefont {Avetian}, \citenamefont {Kruse}, \citenamefont {van Kolck},
  \citenamefont {Maris} \emph {et~al.}}]{Coon:2012ab}%
  \BibitemOpen
  \bibfield  {author} {\bibinfo {author} {\bibfnamefont {S.~A.}\ \bibnamefont
  {Coon}}, \bibinfo {author} {\bibfnamefont {M.~I.}\ \bibnamefont {Avetian}},
  \bibinfo {author} {\bibfnamefont {M.~K.}\ \bibnamefont {Kruse}}, \bibinfo
  {author} {\bibfnamefont {U.}~\bibnamefont {van Kolck}}, \bibinfo {author}
  {\bibfnamefont {P.}~\bibnamefont {Maris}},  \emph {et~al.},\ }\href {\doibase
  10.1103/PhysRevC.86.054002} {\bibfield  {journal} {\bibinfo  {journal} {Phys.
  Rev. C}\ }\textbf {\bibinfo {volume} {86}},\ \bibinfo {pages} {054002}
  (\bibinfo {year} {2012})}\BibitemShut {NoStop}%
\bibitem [{\citenamefont {Furnstahl}\ \emph {et~al.}(2012)\citenamefont
  {Furnstahl}, \citenamefont {Hagen},\ and\ \citenamefont
  {Papenbrock}}]{Furnstahl:2012qg}%
  \BibitemOpen
  \bibfield  {author} {\bibinfo {author} {\bibfnamefont {R.~J.}\ \bibnamefont
  {Furnstahl}}, \bibinfo {author} {\bibfnamefont {G.}~\bibnamefont {Hagen}}, \
  and\ \bibinfo {author} {\bibfnamefont {T.}~\bibnamefont {Papenbrock}},\
  }\href {\doibase 10.1103/PhysRevC.86.031301} {\bibfield  {journal} {\bibinfo
  {journal} {Phys. Rev. C}\ }\textbf {\bibinfo {volume} {86}},\ \bibinfo
  {pages} {031301} (\bibinfo {year} {2012})}\BibitemShut {NoStop}%
\bibitem [{\citenamefont {Hagen}\ \emph {et~al.}(2007)\citenamefont {Hagen}
  \emph {et~al.}}]{Hagen:2007ew}%
  \BibitemOpen
  \bibfield  {author} {\bibinfo {author} {\bibfnamefont {G.}~\bibnamefont
  {Hagen}} \emph {et~al.},\ }\href@noop {} {\bibfield  {journal} {\bibinfo
  {journal} {Phys. Rev. C}\ }\textbf {\bibinfo {volume} {76}},\ \bibinfo
  {pages} {034302} (\bibinfo {year} {2007})}\BibitemShut {NoStop}%
\bibitem [{\citenamefont {Bogner}\ \emph {et~al.}(2008)\citenamefont {Bogner},
  \citenamefont {Furnstahl}, \citenamefont {Maris}, \citenamefont {Perry},
  \citenamefont {Schwenk},\ and\ \citenamefont {Vary}}]{Bogner:2007rx}%
  \BibitemOpen
  \bibfield  {author} {\bibinfo {author} {\bibfnamefont {S.~K.}\ \bibnamefont
  {Bogner}}, \bibinfo {author} {\bibfnamefont {R.~J.}\ \bibnamefont
  {Furnstahl}}, \bibinfo {author} {\bibfnamefont {P.}~\bibnamefont {Maris}},
  \bibinfo {author} {\bibfnamefont {R.~J.}\ \bibnamefont {Perry}}, \bibinfo
  {author} {\bibfnamefont {A.}~\bibnamefont {Schwenk}}, \ and\ \bibinfo
  {author} {\bibfnamefont {J.~P.}\ \bibnamefont {Vary}},\ }\href {\doibase
  10.1016/j.nuclphysa.2007.12.008} {\bibfield  {journal} {\bibinfo  {journal}
  {Nucl. Phys. A}\ }\textbf {\bibinfo {volume} {801}},\ \bibinfo {pages} {21}
  (\bibinfo {year} {2008})}\BibitemShut {NoStop}%
\bibitem [{\citenamefont {Forssen}\ \emph {et~al.}(2008)\citenamefont
  {Forssen}, \citenamefont {Vary}, \citenamefont {Caurier},\ and\ \citenamefont
  {Navratil}}]{Forssen:2008qp}%
  \BibitemOpen
  \bibfield  {author} {\bibinfo {author} {\bibfnamefont {C.}~\bibnamefont
  {Forssen}}, \bibinfo {author} {\bibfnamefont {J.}~\bibnamefont {Vary}},
  \bibinfo {author} {\bibfnamefont {E.}~\bibnamefont {Caurier}}, \ and\
  \bibinfo {author} {\bibfnamefont {P.}~\bibnamefont {Navratil}},\ }\href
  {\doibase 10.1103/PhysRevC.77.024301} {\bibfield  {journal} {\bibinfo
  {journal} {Phys. Rev. C}\ }\textbf {\bibinfo {volume} {77}},\ \bibinfo
  {pages} {024301} (\bibinfo {year} {2008})}\BibitemShut {NoStop}%
\bibitem [{\citenamefont {Maris}\ \emph {et~al.}(2009)\citenamefont {Maris},
  \citenamefont {Vary},\ and\ \citenamefont {Shirokov}}]{Maris:2008ax}%
  \BibitemOpen
  \bibfield  {author} {\bibinfo {author} {\bibfnamefont {P.}~\bibnamefont
  {Maris}}, \bibinfo {author} {\bibfnamefont {J.~P.}\ \bibnamefont {Vary}}, \
  and\ \bibinfo {author} {\bibfnamefont {A.~M.}\ \bibnamefont {Shirokov}},\
  }\href {\doibase 10.1103/PhysRevC.79.014308} {\bibfield  {journal} {\bibinfo
  {journal} {Phys. Rev. C}\ }\textbf {\bibinfo {volume} {79}},\ \bibinfo
  {pages} {014308} (\bibinfo {year} {2009})}\BibitemShut {NoStop}%
\bibitem [{\citenamefont {Roth}(2009)}]{Roth:2009cw}%
  \BibitemOpen
  \bibfield  {author} {\bibinfo {author} {\bibfnamefont {R.}~\bibnamefont
  {Roth}},\ }\href {\doibase 10.1103/PhysRevC.79.064324} {\bibfield  {journal}
  {\bibinfo  {journal} {Phys. Rev. C}\ }\textbf {\bibinfo {volume} {79}},\
  \bibinfo {pages} {064324} (\bibinfo {year} {2009})}\BibitemShut {NoStop}%
\bibitem [{\citenamefont {More}\ \emph {et~al.}(2013)\citenamefont {More},
  \citenamefont {Ekstr{\"o}m}, \citenamefont {Furnstahl}, \citenamefont
  {Hagen},\ and\ \citenamefont {Papenbrock}}]{More:2013rma}%
  \BibitemOpen
  \bibfield  {author} {\bibinfo {author} {\bibfnamefont {S.}~\bibnamefont
  {More}}, \bibinfo {author} {\bibfnamefont {A.}~\bibnamefont {Ekstr{\"o}m}},
  \bibinfo {author} {\bibfnamefont {R.}~\bibnamefont {Furnstahl}}, \bibinfo
  {author} {\bibfnamefont {G.}~\bibnamefont {Hagen}}, \ and\ \bibinfo {author}
  {\bibfnamefont {T.}~\bibnamefont {Papenbrock}},\ }\href {\doibase
  10.1103/PhysRevC.87.044326} {\bibfield  {journal} {\bibinfo  {journal} {Phys.
  Rev. C}\ }\textbf {\bibinfo {volume} {87}},\ \bibinfo {pages} {044326}
  (\bibinfo {year} {2013})}\BibitemShut {NoStop}%
\bibitem [{\citenamefont {Soma}\ \emph {et~al.}(2013)\citenamefont {Soma},
  \citenamefont {Barbieri},\ and\ \citenamefont {Duguet}}]{Soma:2012zd}%
  \BibitemOpen
  \bibfield  {author} {\bibinfo {author} {\bibfnamefont {V.}~\bibnamefont
  {Soma}}, \bibinfo {author} {\bibfnamefont {C.}~\bibnamefont {Barbieri}}, \
  and\ \bibinfo {author} {\bibfnamefont {T.}~\bibnamefont {Duguet}},\ }\href
  {\doibase 10.1103/PhysRevC.87.011303} {\bibfield  {journal} {\bibinfo
  {journal} {Phys. Rev. C}\ }\textbf {\bibinfo {volume} {87}},\ \bibinfo
  {pages} {011303} (\bibinfo {year} {2013})}\BibitemShut {NoStop}%
\bibitem [{\citenamefont {Hergert}\ \emph {et~al.}(2013)\citenamefont
  {Hergert}, \citenamefont {Bogner}, \citenamefont {Binder}, \citenamefont
  {Calci}, \citenamefont {Langhammer}, \citenamefont {Roth},\ and\
  \citenamefont {Schwenk}}]{Hergert:2012}%
  \BibitemOpen
  \bibfield  {author} {\bibinfo {author} {\bibfnamefont {H.}~\bibnamefont
  {Hergert}}, \bibinfo {author} {\bibfnamefont {S.~K.}\ \bibnamefont {Bogner}},
  \bibinfo {author} {\bibfnamefont {S.}~\bibnamefont {Binder}}, \bibinfo
  {author} {\bibfnamefont {A.}~\bibnamefont {Calci}}, \bibinfo {author}
  {\bibfnamefont {J.}~\bibnamefont {Langhammer}}, \bibinfo {author}
  {\bibfnamefont {R.}~\bibnamefont {Roth}}, \ and\ \bibinfo {author}
  {\bibfnamefont {A.}~\bibnamefont {Schwenk}},\ }\href {\doibase
  10.1103/PhysRevC.87.034307} {\bibfield  {journal} {\bibinfo  {journal} {Phys.
  Rev. C}\ }\textbf {\bibinfo {volume} {87}},\ \bibinfo {pages} {034307}
  (\bibinfo {year} {2013})}\BibitemShut {NoStop}%
\bibitem [{\citenamefont {Jurgenson}\ \emph {et~al.}(2013)\citenamefont
  {Jurgenson}, \citenamefont {Maris}, \citenamefont {Furnstahl}, \citenamefont
  {Navr\'atil}, \citenamefont {Ormand},\ and\ \citenamefont
  {Vary}}]{Jurgenson:2013}%
  \BibitemOpen
  \bibfield  {author} {\bibinfo {author} {\bibfnamefont {E.~D.}\ \bibnamefont
  {Jurgenson}}, \bibinfo {author} {\bibfnamefont {P.}~\bibnamefont {Maris}},
  \bibinfo {author} {\bibfnamefont {R.~J.}\ \bibnamefont {Furnstahl}}, \bibinfo
  {author} {\bibfnamefont {P.}~\bibnamefont {Navr\'atil}}, \bibinfo {author}
  {\bibfnamefont {W.~E.}\ \bibnamefont {Ormand}}, \ and\ \bibinfo {author}
  {\bibfnamefont {J.~P.}\ \bibnamefont {Vary}},\ }\href {\doibase
  10.1103/PhysRevC.87.054312} {\bibfield  {journal} {\bibinfo  {journal} {Phys.
  Rev. C}\ }\textbf {\bibinfo {volume} {87}},\ \bibinfo {pages} {054312}
  (\bibinfo {year} {2013})}\BibitemShut {NoStop}%
\bibitem [{\citenamefont {{S\"a\"af}}\ and\ \citenamefont
  {Forss{\'e}n}(2014)}]{Saaf:2013asa}%
  \BibitemOpen
  \bibfield  {author} {\bibinfo {author} {\bibfnamefont {D.}~\bibnamefont
  {{S\"a\"af}}}\ and\ \bibinfo {author} {\bibfnamefont {C.}~\bibnamefont
  {Forss{\'e}n}},\ }\href {\doibase 10.1103/PhysRevC.89.011303} {\bibfield
  {journal} {\bibinfo  {journal} {Phys. Rev. C}\ }\textbf {\bibinfo {volume}
  {89}},\ \bibinfo {pages} {011303} (\bibinfo {year} {2014})}\BibitemShut
  {NoStop}%
\bibitem [{\citenamefont {Roth}\ \emph {et~al.}(2013)\citenamefont {Roth},
  \citenamefont {Calci}, \citenamefont {Langhammer},\ and\ \citenamefont
  {Binder}}]{Roth:2013fqa}%
  \BibitemOpen
  \bibfield  {author} {\bibinfo {author} {\bibfnamefont {R.}~\bibnamefont
  {Roth}}, \bibinfo {author} {\bibfnamefont {A.}~\bibnamefont {Calci}},
  \bibinfo {author} {\bibfnamefont {J.}~\bibnamefont {Langhammer}}, \ and\
  \bibinfo {author} {\bibfnamefont {S.}~\bibnamefont {Binder}},\ }\href@noop {}
  {(\bibinfo {year} {2013})},\ \Eprint {http://arxiv.org/abs/1311.3563}
  {arXiv:1311.3563 [nucl-th]} \BibitemShut {NoStop}%
\bibitem [{\citenamefont {L{\"u}scher}(1986)}]{Luscher:1985dn}%
  \BibitemOpen
  \bibfield  {author} {\bibinfo {author} {\bibfnamefont {M.}~\bibnamefont
  {L{\"u}scher}},\ }\href {\doibase 10.1007/BF01211589} {\bibfield  {journal}
  {\bibinfo  {journal} {Commun. Math. Phys.}\ }\textbf {\bibinfo {volume}
  {104}},\ \bibinfo {pages} {177} (\bibinfo {year} {1986})}\BibitemShut
  {NoStop}%
\bibitem [{\citenamefont {Lee}\ and\ \citenamefont {Pine}(2011)}]{Lee:2010km}%
  \BibitemOpen
  \bibfield  {author} {\bibinfo {author} {\bibfnamefont {D.}~\bibnamefont
  {Lee}}\ and\ \bibinfo {author} {\bibfnamefont {M.}~\bibnamefont {Pine}},\
  }\href {\doibase 10.1140/epja/i2011-11041-4} {\bibfield  {journal} {\bibinfo
  {journal} {Eur. Phys. J. A}\ }\textbf {\bibinfo {volume} {47}},\ \bibinfo
  {pages} {41} (\bibinfo {year} {2011})}\BibitemShut {NoStop}%
\bibitem [{\citenamefont {Davoudi}\ and\ \citenamefont
  {Savage}(2011)}]{Zohreh:2011}%
  \BibitemOpen
  \bibfield  {author} {\bibinfo {author} {\bibfnamefont {Z.}~\bibnamefont
  {Davoudi}}\ and\ \bibinfo {author} {\bibfnamefont {M.~J.}\ \bibnamefont
  {Savage}},\ }\href {\doibase 10.1103/PhysRevD.84.114502} {\bibfield
  {journal} {\bibinfo  {journal} {Phys. Rev. D}\ }\textbf {\bibinfo {volume}
  {84}},\ \bibinfo {pages} {114502} (\bibinfo {year} {2011})}\BibitemShut
  {NoStop}%
\bibitem [{\citenamefont {Koenig}\ \emph {et~al.}(2011)\citenamefont {Koenig},
  \citenamefont {Lee},\ and\ \citenamefont {Hammer}}]{Konig:2011nz}%
  \BibitemOpen
  \bibfield  {author} {\bibinfo {author} {\bibfnamefont {S.}~\bibnamefont
  {Koenig}}, \bibinfo {author} {\bibfnamefont {D.}~\bibnamefont {Lee}}, \ and\
  \bibinfo {author} {\bibfnamefont {H.-W.}\ \bibnamefont {Hammer}},\ }\href
  {\doibase 10.1103/PhysRevLett.107.112001} {\bibfield  {journal} {\bibinfo
  {journal} {Phys. Rev. Lett.}\ }\textbf {\bibinfo {volume} {107}},\ \bibinfo
  {pages} {112001} (\bibinfo {year} {2011})}\BibitemShut {NoStop}%
\bibitem [{\citenamefont {{Pine}}\ and\ \citenamefont
  {{Lee}}(2013)}]{Pine:2013}%
  \BibitemOpen
  \bibfield  {author} {\bibinfo {author} {\bibfnamefont {M.}~\bibnamefont
  {{Pine}}}\ and\ \bibinfo {author} {\bibfnamefont {D.}~\bibnamefont {{Lee}}},\
  }\href {\doibase 10.1016/j.aop.2012.12.009} {\bibfield  {journal} {\bibinfo
  {journal} {Annals of Physics}\ }\textbf {\bibinfo {volume} {331}},\ \bibinfo
  {pages} {24} (\bibinfo {year} {2013})}\BibitemShut {NoStop}%
\bibitem [{\citenamefont {Brice\~no}\ \emph {et~al.}(2013)\citenamefont
  {Brice\~no}, \citenamefont {Davoudi}, \citenamefont {Luu},\ and\
  \citenamefont {Savage}}]{Briceno:2013}%
  \BibitemOpen
  \bibfield  {author} {\bibinfo {author} {\bibfnamefont {R.~A.}\ \bibnamefont
  {Brice\~no}}, \bibinfo {author} {\bibfnamefont {Z.}~\bibnamefont {Davoudi}},
  \bibinfo {author} {\bibfnamefont {T.~C.}\ \bibnamefont {Luu}}, \ and\
  \bibinfo {author} {\bibfnamefont {M.~J.}\ \bibnamefont {Savage}},\ }\href
  {\doibase 10.1103/PhysRevD.88.114507} {\bibfield  {journal} {\bibinfo
  {journal} {Phys. Rev. D}\ }\textbf {\bibinfo {volume} {88}},\ \bibinfo
  {pages} {114507} (\bibinfo {year} {2013})}\BibitemShut {NoStop}%
\bibitem [{\citenamefont {Bulgac}\ and\ \citenamefont
  {Forbes}(2013)}]{Bulgac:2013mz}%
  \BibitemOpen
  \bibfield  {author} {\bibinfo {author} {\bibfnamefont {A.}~\bibnamefont
  {Bulgac}}\ and\ \bibinfo {author} {\bibfnamefont {M.~M.}\ \bibnamefont
  {Forbes}},\ }\href {\doibase 10.1103/PhysRevC.87.051301} {\bibfield
  {journal} {\bibinfo  {journal} {Phys. Rev. C}\ }\textbf {\bibinfo {volume}
  {87}},\ \bibinfo {pages} {051301} (\bibinfo {year} {2013})}\BibitemShut
  {NoStop}%
\bibitem [{\citenamefont {Djajaputra}\ and\ \citenamefont
  {Cooper}(2000)}]{Djajaputra:2000aa}%
  \BibitemOpen
  \bibfield  {author} {\bibinfo {author} {\bibfnamefont {D.}~\bibnamefont
  {Djajaputra}}\ and\ \bibinfo {author} {\bibfnamefont {B.~R.}\ \bibnamefont
  {Cooper}},\ }\href {http://stacks.iop.org/0143-0807/21/i=3/a=309} {\bibfield
  {journal} {\bibinfo  {journal} {Eur. J. Phys.}\ }\textbf
  {\bibinfo {volume} {21}},\ \bibinfo {pages} {261} (\bibinfo {year}
  {2000})}\BibitemShut {NoStop}%
\bibitem [{\citenamefont {Taylor}(2006)}]{taylor2006scattering}%
  \BibitemOpen
  \bibfield  {author} {\bibinfo {author} {\bibfnamefont {J.~R.}\ \bibnamefont
  {Taylor}},\ }\href@noop {} {\emph {\bibinfo {title} {Scattering Theory: The
  Quantum Theory of Nonrelativistic Collisions}}}\ (\bibinfo  {publisher}
  {Dover},\ \bibinfo {address} {New York},\ \bibinfo {year} {2006})\BibitemShut
  {NoStop}%
\bibitem [{\citenamefont {Newton}(2002)}]{newton2002scattering}%
  \BibitemOpen
  \bibfield  {author} {\bibinfo {author} {\bibfnamefont {R.~G.}\ \bibnamefont
  {Newton}},\ }\href@noop {} {\emph {\bibinfo {title} {Scattering theory of
  waves and particles}}}\ (\bibinfo  {publisher} {Dover},\ \bibinfo {address}
  {New York},\ \bibinfo {year} {2002})\BibitemShut {NoStop}%
\bibitem [{\citenamefont {Wu}\ and\ \citenamefont
  {Ohmura}(2011)}]{wu2011scattering}%
  \BibitemOpen
  \bibfield  {author} {\bibinfo {author} {\bibfnamefont {T.-Y.}\ \bibnamefont
  {Wu}}\ and\ \bibinfo {author} {\bibfnamefont {T.}~\bibnamefont {Ohmura}},\
  }\href@noop {} {\emph {\bibinfo {title} {Quantum Theory of Scattering}}}\
  (\bibinfo  {publisher} {Dover},\ \bibinfo {address} {New York},\ \bibinfo
  {year} {2011})\BibitemShut {NoStop}%
\bibitem [{\citenamefont {Phillips}\ \emph {et~al.}(2000)\citenamefont
  {Phillips}, \citenamefont {Rupak},\ and\ \citenamefont
  {Savage}}]{Phillips:1999hh}%
  \BibitemOpen
  \bibfield  {author} {\bibinfo {author} {\bibfnamefont {D.~R.}\ \bibnamefont
  {Phillips}}, \bibinfo {author} {\bibfnamefont {G.}~\bibnamefont {Rupak}}, \
  and\ \bibinfo {author} {\bibfnamefont {M.~J.}\ \bibnamefont {Savage}},\
  }\href {\doibase 10.1016/S0370-2693(99)01496-3} {\bibfield  {journal}
  {\bibinfo  {journal} {Phys. Lett. B}\ }\textbf {\bibinfo {volume} {473}},\
  \bibinfo {pages} {209} (\bibinfo {year} {2000})}\BibitemShut {NoStop}%
\bibitem [{\citenamefont {Entem}\ and\ \citenamefont
  {Machleidt}(2003)}]{Entem:2003ft}%
  \BibitemOpen
  \bibfield  {author} {\bibinfo {author} {\bibfnamefont {D.~R.}\ \bibnamefont
  {Entem}}\ and\ \bibinfo {author} {\bibfnamefont {R.}~\bibnamefont
  {Machleidt}},\ }\href@noop {} {\bibfield  {journal} {\bibinfo  {journal}
  {Phys. Rev. C}\ }\textbf {\bibinfo {volume} {68}},\ \bibinfo {pages} {041001}
  (\bibinfo {year} {2003})}\BibitemShut {NoStop}%
\bibitem [{\citenamefont {Dea{\~n}o}\ \emph {et~al.}(2013)\citenamefont
  {Dea{\~n}o}, \citenamefont {Huertas},\ and\ \citenamefont
  {Marcell{\'a}n}}]{Deano2013}%
  \BibitemOpen
  \bibfield  {author} {\bibinfo {author} {\bibfnamefont {A.}~\bibnamefont
  {Dea{\~n}o}}, \bibinfo {author} {\bibfnamefont {E.~J.}\ \bibnamefont
  {Huertas}}, \ and\ \bibinfo {author} {\bibfnamefont {F.}~\bibnamefont
  {Marcell{\'a}n}},\ }\href {\doibase 10.1016/j.jmaa.2013.02.039} {\bibfield
  {journal} {\bibinfo  {journal} {J. Math. Anal. Appl.}\ }\textbf {\bibinfo
  {volume} {403}},\ \bibinfo {pages} {477} (\bibinfo {year}
  {2013})}\BibitemShut {NoStop}%
\bibitem [{\citenamefont {Abramowitz}\ and\ \citenamefont
  {Stegun}(1972)}]{abramowitz1964}%
  \BibitemOpen
  \bibfield  {author} {\bibinfo {author} {\bibfnamefont {M.}~\bibnamefont
  {Abramowitz}}\ and\ \bibinfo {author} {\bibfnamefont {I.~A.}\ \bibnamefont
  {Stegun}},\ }\href@noop {} {\emph {\bibinfo {title} {Handbook of Mathematical
  Functions}}}\ (\bibinfo  {publisher} {Dover},\ \bibinfo {address} {New
  York},\ \bibinfo {year} {1972})\BibitemShut {NoStop}%
\bibitem [{\citenamefont {Machleidt}\ and\ \citenamefont
  {Entem}(2011)}]{Machleidt:2011zz}%
  \BibitemOpen
  \bibfield  {author} {\bibinfo {author} {\bibfnamefont {R.}~\bibnamefont
  {Machleidt}}\ and\ \bibinfo {author} {\bibfnamefont {D.}~\bibnamefont
  {Entem}},\ }\href {\doibase 10.1016/j.physrep.2011.02.001} {\bibfield
  {journal} {\bibinfo  {journal} {Phys. Rept.}\ }\textbf {\bibinfo {volume}
  {503}},\ \bibinfo {pages} {1} (\bibinfo {year} {2011})}\BibitemShut {NoStop}%
\bibitem [{\citenamefont {Arteca}\ \emph {et~al.}(1984)\citenamefont {Arteca},
  \citenamefont {Fern{\'a}ndez},\ and\ \citenamefont {Castro}}]{Arteca1984}%
  \BibitemOpen
  \bibfield  {author} {\bibinfo {author} {\bibfnamefont {G.~A.}\ \bibnamefont
  {Arteca}}, \bibinfo {author} {\bibfnamefont {F.~M.}\ \bibnamefont
  {Fern{\'a}ndez}}, \ and\ \bibinfo {author} {\bibfnamefont {E.~A.}\
  \bibnamefont {Castro}},\ }\href@noop {} {\bibfield  {journal} {\bibinfo
  {journal} {J. of Chem. Phys.}\ }\textbf {\bibinfo {volume} {80}},\ \bibinfo
  {pages} {1569} (\bibinfo {year} {1984})}\BibitemShut {NoStop}%
\bibitem [{\citenamefont {Fern{\'a}ndez}\ and\ \citenamefont
  {Castro}(1981)}]{Fernandez1981}%
  \BibitemOpen
  \bibfield  {author} {\bibinfo {author} {\bibfnamefont {F.~M.}\ \bibnamefont
  {Fern{\'a}ndez}}\ and\ \bibinfo {author} {\bibfnamefont {E.~A.}\ \bibnamefont
  {Castro}},\ }\href {\doibase 10.1002/qua.560190406} {\bibfield  {journal}
  {\bibinfo  {journal} {Int. J. of Quantum Chem.}\ }\textbf {\bibinfo {volume}
  {19}},\ \bibinfo {pages} {521} (\bibinfo {year} {1981})}\BibitemShut
  {NoStop}%
\bibitem [{\citenamefont {Stetcu}\ \emph {et~al.}(2005)\citenamefont {Stetcu},
  \citenamefont {Barrett}, \citenamefont {Navratil},\ and\ \citenamefont
  {Vary}}]{Stetcu:2004wh}%
  \BibitemOpen
  \bibfield  {author} {\bibinfo {author} {\bibfnamefont {I.}~\bibnamefont
  {Stetcu}}, \bibinfo {author} {\bibfnamefont {B.~R.}\ \bibnamefont {Barrett}},
  \bibinfo {author} {\bibfnamefont {P.}~\bibnamefont {Navratil}}, \ and\
  \bibinfo {author} {\bibfnamefont {J.~P.}\ \bibnamefont {Vary}},\ }\href
  {\doibase 10.1103/PhysRevC.71.044325} {\bibfield  {journal} {\bibinfo
  {journal} {Phys. Rev. C}\ }\textbf {\bibinfo {volume} {71}},\ \bibinfo
  {pages} {044325} (\bibinfo {year} {2005})}\BibitemShut {NoStop}%
\end{thebibliography}%

\end{document}